\documentclass[5p,authoryear]{elsarticle} 
\usepackage{lineno,hyperref}
\usepackage{amsmath}
\usepackage{amssymb}
\usepackage{verbatim}
\usepackage{multirow}
\usepackage{color,graphicx}
\usepackage{arydshln}
\usepackage{hyperref}
\usepackage{url}
\usepackage{footnote}
\usepackage{bm}
\usepackage{cite}
\usepackage{subfigure}
\usepackage{xcolor}
\usepackage{ulem}

\usepackage{nicematrix}
\usepackage{ulem}
\usepackage{latexsym}

\modulolinenumbers[5]

\providecommand{\Leireftb}[1]{Table~\ref{#1}}
\providecommand{\Leireffig}[1]{Fig.~\ref{#1}}

\definecolor{Mycolor1}{HTML}{036564}
\definecolor{Mycolor2}{HTML}{DE7D2C}

\usepackage{enumitem}

\journal{Medical Image Analysis}

\begin{document}


\begin{frontmatter}

\title{Multi-Modality Cardiac Image Computing: A Survey}

\author[label1]{Lei Li} 
\ead[url]{lei.li@eng.ox.ac.uk}
\author[label3]{Wangbin Ding} 
\author[label3]{Liqun Huang} 
\author[label2]{Xiahai Zhuang*} 
\author[label1]{Vicente Grau*} 

\cortext[cor]{Both are co-senior authors and contribute equally.}
 
\address[label1]{Department of Engineering Science, University of Oxford, Oxford, UK}
\address[label2]{School of Data Science, Fudan University, Shanghai, China}
\address[label3]{College of Physics and Information Engineering, Fuzhou University, Fuzhou, China}


\begin{abstract}
Multi-modality cardiac imaging plays a key role in the management of patients with cardiovascular diseases.
It allows a combination of complementary anatomical, morphological and functional information, increases diagnosis accuracy, and improves the efficacy of cardiovascular interventions and clinical outcomes.
Fully-automated processing and quantitative analysis of multi-modality cardiac images could have a direct impact on clinical research and evidence-based patient management.
However, these require overcoming significant challenges including inter-modality misalignment and finding optimal methods to integrate information from different modalities.

This paper aims to provide a comprehensive review of multi-modality imaging in cardiology, the computing methods, the validation strategies, the related clinical workflows and future perspectives.
For the computing methodologies, we have a favored focus on the three tasks, i.e., registration, fusion and segmentation, which generally involve multi-modality imaging data, \textit{either combining information from different modalities or transferring information across modalities}. 
The review highlights that multi-modality cardiac imaging data has the potential of wide applicability in the clinic, such as trans-aortic valve implantation guidance, myocardial viability assessment, and catheter ablation therapy and its patient selection.
Nevertheless, many challenges remain unsolved, such as missing modality, combination of imaging and non-imaging data, and uniform analysis and representation of different modalities.
There is also work to do in defining how the well-developed techniques fit in clinical workflows and how much additional and relevant information they introduce.
These problems are likely to continue to be an active field of research and the questions to be answered in the future.
\end{abstract}


\end{frontmatter}


\section{Introduction}

\begin{figure*}[t]\center
 \includegraphics[width=0.98\textwidth]{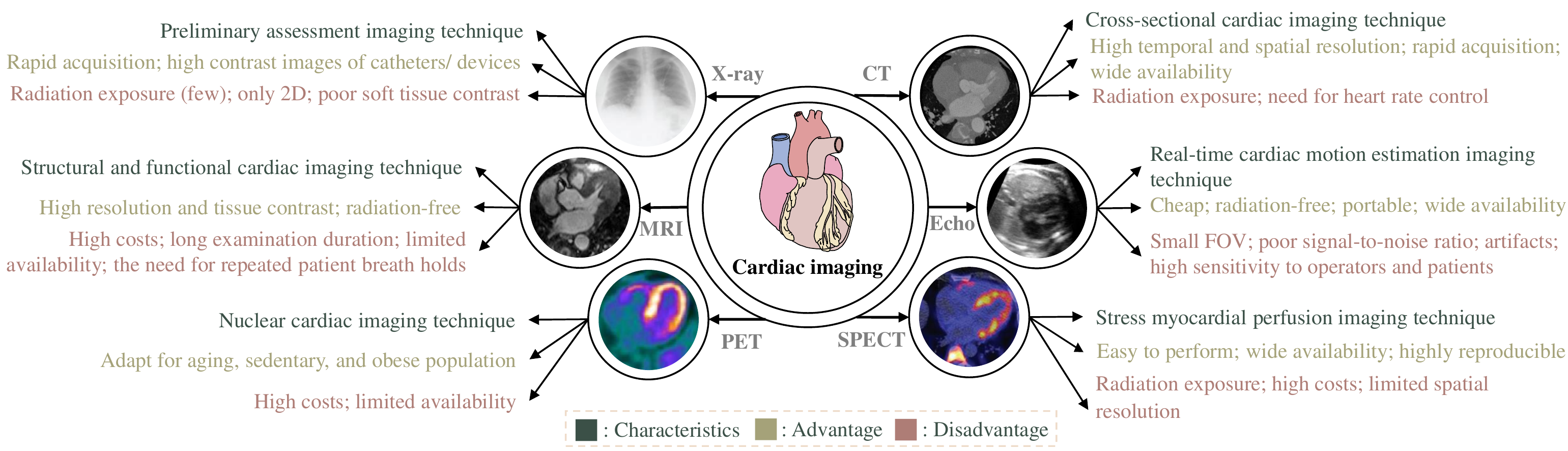}\\[-2ex] 
   \caption{Examples of cardiac imaging techniques from different modalities with their characteristics, advantages and disadvantages. 
   CT: computed tomography; MRI: magnetic resonance imaging; Echo: echocardiography; PET: positron emission tomography; SPECT: single-photon emission computed tomography.
   Here, X-ray, PET and SPECT images adapted from \citet{journal/JCARS/kholiavchenko2020}, \citet{journal/HFR/quail2017} and \citet{journal/CFR/klaassen2020} with permission, respectively.}
\label{fig:intro:MM-cardiac images}
\end{figure*}

Cardiovascular diseases (CVDs) remain one of the leading causes of mortality worldwide.
Non-invasive cardiac imaging techniques can provide anatomical and functional information on health and pathology, and have been developed and employed extensively for the diagnosis and treatment of CVDs.
Several imaging modalities are used for cardiac assessment, including X-ray, computed tomography (CT), magnetic resonance imaging (MRI), echocardiography (Echo), positron emission tomography (PET), and single-photon emission computed tomography (SPECT) \citep{journal/CDT/laczay2021}.
As \Leireffig{fig:intro:MM-cardiac images} shows, each imaging modality manifests particular information, and a single modality may provide insufficient and/or ambiguous information regarding the condition of the heart that is a nonrigid and dynamic structure.
Combining the complementary information from multi-modality cardiac images can be beneficial in establishing a more accurate diagnosis or assisting clinicians in the treatment management of CVDs \citep{journal/EHJ/valsangiacomo2012,journal/CI/olsen2016,journal/MedIA/puyol2021}.

However, automated cardiac image computing is particularly challenging on account of normal or pathologic deformation of the heart during the respiratory and cardiac cycle.  
Furthermore, there exist misalignment and intensity variations among different modalities, and the imaging phases, field-of-views (FOVs), and resolution of different acquisitions can vary significantly.
Several research methods have been developed to process multi-modality cardiac images, including \textit{registration}, \textit{fusion}, and \textit{segmentation} of multi-modality images.
With multi-modality images, image registration aims to align the available measurements (shape, function, pathology, etc.) into the same anatomical reference \citep{journal/TPAMI/Zhuang2019}. 
Image fusion intends to integrate several imaging modalities in an optimal way for visualisation or decision-making. 
Image segmentation attempts to extract target regions by combining the complementary information from several imaging modalities.
The three tasks exhibit overlap as well as distinct challenges.
Specifically, their common issues are related to various sources of motion (patient, respiratory, and cardiac cycle).
Unique challenges for registration and fusion may arise from the lack of a gold standard, which disables most current supervised learning-based models.
The challenges associated with segmentation could originate from specific target regions, which can have large shape variations (e.g. the left atrium (LA)), can be small and/or diffuse (such as fibrotic regions and scars), or thin compared to image resolution (such as LA wall and coronary arteries).
The proper selection of fusion techniques is complicated, as they should be easily interpretable in order to obtain traceable results.
Also, the algorithms must satisfy several requirements, such as robustness to different acquisition conditions or tolerable computation times in a real-time system.

\begin{figure}[t]\center
    \includegraphics[width=0.49\textwidth]{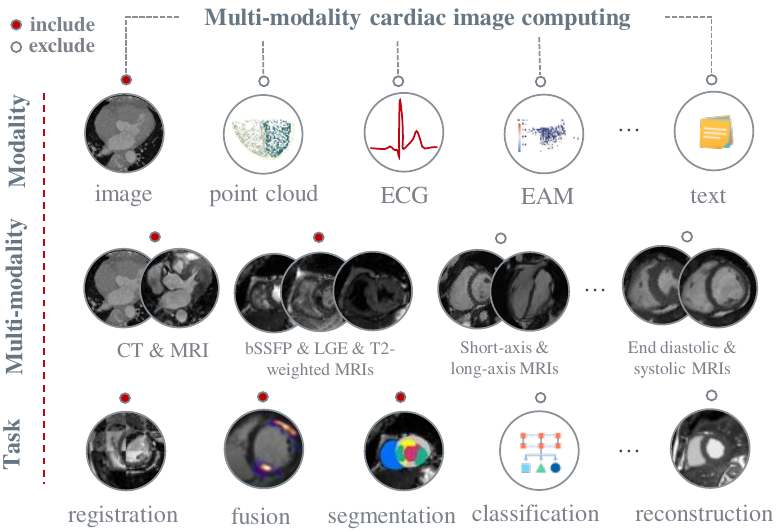}
  \caption{The scope of this review.
  Here, the solid and hollow circles at the top of each category indicate inclusion and exclusion, respectively.}
\label{fig:intro:review_scope}
\end{figure}

\begin{table} [t] \center
    \caption{Search engines and expressions used to identify potential papers for review. 
     }
\label{tb:intro:search item}
{\small
\begin{tabular}{p{0.75cm}|p{7.2cm}}
\hline
Engine                 & Google scholar, PubMed, IEEE-Xplore, and Citeseer\\
\hline
\multirow{11}{*}{Term}  &``Multi(-)modality/ modal/ sequence/ source" or  ``cross(-)modality/ modal/ sequence" \textbf{and} \\ \cline{2-2}
                       & ``X-ray" or ``computed tomography/ CT" or ``magnetic resonance/ MR(I)" or ``ultrasound/ US" or ``echocardiography/ Echo" or ``positron emission tomography/ PET" or ``single-photon emission computed tomography/ SPECT" or ``angiography" or ``angiogram"  \textbf{and} \\ \cline{2-2} 
                      ~& ``Cardiac" or ``heart" or ``atri$^*$" or ``ventricl$^*$" or ``myocardi$^*$" or ``aort$^*$" or ``coronary artery" \textbf{and} \\ \cline{2-2} 
                      ~& ``Regist$^*$" or ``fus$^*$" or ``integrat$^*$" or ``combin$^*$" or ``hybrid" or ``overlay" or ``segment$^*$" or ``domain adaptation"\\
\hline
\end{tabular} }\\
\end{table}

\begin{figure}[t]\center
    \includegraphics[width=0.42\textwidth]{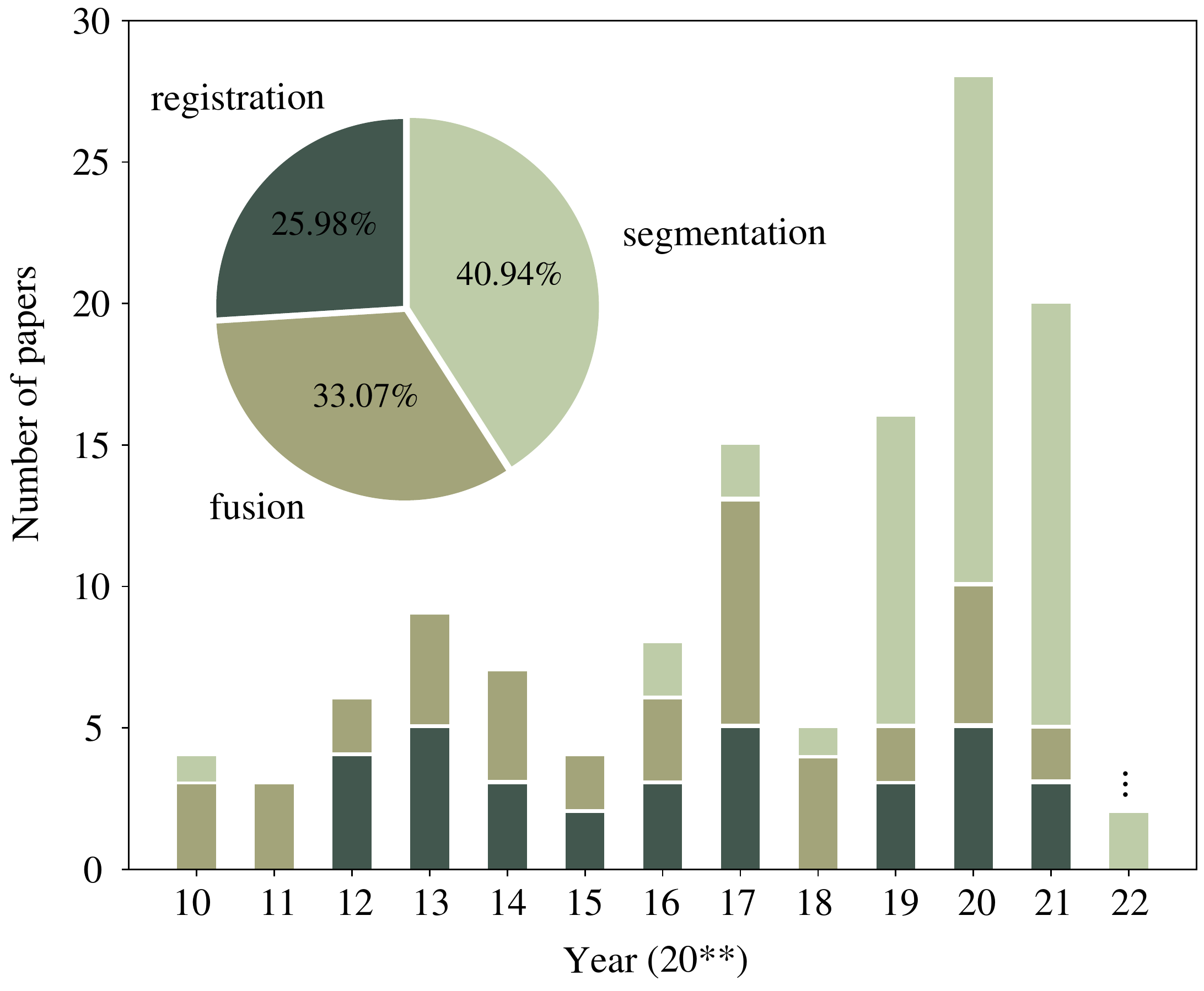}
   \caption{The distributions of reviewed papers of multi-modality cardiac image computing per year and task.}
\label{fig:intro:review_distribution}
\end{figure}

\begin{table*} [!t] \center
    \caption{Summary of the related representative review works. 
    ML: machine learning; CAD: coronary artery disease; WH: whole heart; LA: left atrium; RV: right ventricle; DL: deep learning; 
    TMI: IEEE Transactions on Medical Imaging; 
    MedIA: Medical Image Analysis; 
    CRP: Cardiology Research and Practice;
    CVIU: Computer Vision and Image Understanding; 
    JBR: Journal of Biomedical Research; 
    JHE: Journal of Healthcare Engineering; 
    EIJ: Egyptian Informatics Journal; MAGMA: Magnetic Resonance Materials in Physics, Biology and Medicine; 
    FCM: Frontiers in Cardiovascular Medicine; 
    IPADCD: Image Processing for Automated Diagnosis of Cardiac Diseases; 
    EP: EP Europace; 
    CET: Cardiovascular Engineering and Technology;
    IET: IET Image Processing;
    FIP: Frontiers in Physiology;
    CCRR: Current Cardiovascular Risk Reports;
    EJNMMI: European journal of nuclear medicine and molecular imaging;
    JT: Jurnal Teknologi;
    CMIG: Computerized Medical Imaging and Graphics;
    JNM: Journal of Nuclear Medicine;
    CCIR: Current cardiovascular imaging reports.
    Note that here ``single modality" refers to the methods designed for a single modality but they may be evaluated on several different modalities, separately.
     }
    \label{tb:intro:review paper}
    {\small 
    \scalebox{0.9}
    {
    \begin{tabular}{p{4cm}|p{1.2cm}p{7cm}p{6.2cm}}
    \hline
    Source	   & Venue &  Scope  & Inclusion \\
    \hline
    \citet{journal/TMI/makela2002}      & TMI      & cardiac image registration            & partially includes multi-modality \\
    \citet{journal/EJNMMI/slomka2009}   & EJNMMI   & multi-modality image registration     & software based; partially includes heart \\
    \citet{journal/CVIU/tavakoli2013}   & CVIU     & cardiac image segmentation and registration & single modality; shape-based methods  \\   
    \citet{journal/JT/ming2015}         & JT       & cardiac US and CT registration        & US and CT registration \\    
    \citet{journal/EIJ/el2016}          & EIJ      & medical image registration and fusion & partially includes multi-modality or heart  \\
    \citet{journal/CRP/khalil2018}      & CRP      & cardiac image registration            & partially includes multi-modality \\
    \citet{journal/JNM/rischpler2013}   & JNM      & cardiac PET and MRI fusion            & clinical review; PET/ MRI fusion  \\
    \citet{journal/JBR/piccinelli2013}  & JBR      & multi-modality cardiac image fusion   & clinical review; focus on CAD  \\
    \citet{journal/IPADCD/chauhan2021}  & IPADCD   & cardiac image fusion                  & partially includes multi-modality  \\   
    \citet{journal/MedIA/petitjean2011} & MedIA    & cardiac short-axis MRI segmentation   & single modality \\
    \citet{journal/JHE/zhuang2013}      & JHE      & WH segmentation                       & single modality; conventional methods \\
    \citet{journal/MAGMA/peng2016}      & MAGMA    & cardiac MRI segmentation              & single modality \\
    \citet{journal/EP/pontecorboli2017} & EP       & fibrosis segmentation from LGE MRI    & single modality; thresholding methods \\
    \citet{journal/FCM/jamart2020}      & FCM      & LA cavity segmentation from LGE MRI   & single modality; DL-based methods \\
    \citet{journal/CET/habijan2020}     & CET      & WH and chamber segmentation           & single-modality \\ 
    \citet{journal/FCM/chen2020}        & FCM      & DL-based cardiac segmentation         & single modality; DL-based methods  \\    
    \citet{journal/IET/ammari2021}      & IET      & RV segmentation from short-axis MRI   & single modality; RV segmentation \\
    \citet{journal/FIP/wu2021}          & FIP      & scar/ fibrosis segmentation from MRI  & single modality; scar/ fibrosis segmentation \\
    \citet{journal/CMIG/jia2021}        & CMIG     & learning-based vessel segmentation    & single modality; learning-based methods \\   
    \citet{journal/CCRR/kwan2021}       & CCRR     & DL-based cardiac image segmentation   & partially includes multi-modality \\
    \citet{journal/MedIA/li2022b}       & MedIA    & LA-related segmentation from LGE MRI  & single modality; LA-related segmentation \\ 
    \hline
\end{tabular}}
}
\end{table*}


\subsection{Study inclusion and literature search}
In this work, we aim to provide readers with a survey of the state-of-the-art in multi-modality cardiac image computing techniques and the related literature.
\Leireffig{fig:intro:review_scope} illustrates the review scope of this survey.
Note that in the computer vision area the modality might refer to text, audio and images, but in this work we restrict modality to imaging techniques.
\textit{We consider multi-modality cardiac computing to include studies that involve diverse modalities, either combining or transferring information between cardiac image modalities.}  
In the deep learning-based framework, different modalities may be present during training or more critically at the test phase at the same time.
Therefore, we also include cardiac studies aiming at image translation or cross-modality domain adaptation, although different modality images may only appear separately in the training and test stages.
Note that works that only test their methods in different cardiac imaging modalities for evaluation will not be included in this review.
As for the cardiac computing tasks, we only consider the three major tasks at the current state of the art, i.e., registration, fusion, and segmentation.
Classification is also a primary aim, but current classification works in the literature mainly integrate features from different cardiac signals, such as the electrocardiogram (ECG) and phonocardiogram (PCG), rather than imaging techniques \citep{journal/PM/kalidas2016,journal/BSPC/li2021}. 
There exist several cardiac image reconstruction works with the assistance of other modalities, mainly for cardiac motion correction  \citep{journal/PTRSA/polycarpou2021,journal/TBME/sang2021}, which however is under-researched yet.


To ensure comprehensive coverage, we have screened publications mainly from the last 12 years (2010-2021) related to this topic.
Our main sources of reference were Internet searches using engines, including Google Scholar, PubMed, IEEE-Xplore, Connected Papers, and Citeseer.
To cover as many related works as possible, flexible search terms have been employed when using these search engines, as summarized in \Leireftb{tb:intro:search item}.
Both peer-reviewed journal papers and conference papers were included here.
In the way described above, we have collected a comprehensive library of 147 papers.
\Leireffig{fig:intro:review_distribution} presents the distributions of papers in multi-modality cardiac image computing per year/task.
Note that we picked the most detailed and representative ones for this review when we encountered several papers from the same authors about the same subject.
Moreover, as multi-modality cardiac image fusion works are more likely to be driven by a clinical objective rather than technical development, we only select several representative studies here to cover as many modality combinations as possible. 
Also, the work of fusion or segmentation may involve a registration step, but we usually only regard this work as fusion or segmentation when counting the number of papers, since these are the final aims.

\subsection{Existing survey from literature}

\Leireftb{tb:intro:review paper} lists current review papers related to the topic, i.e., multi-modality cardiac image computing.
One can see that the scopes of current review works are different from ours though with partial overlaps.
For example, current review works on cardiac image registration only summarized conventional registration methods \citep{journal/TMI/makela2002,journal/CVIU/tavakoli2013}, only revolved around specific modalities \citep{journal/JT/ming2015}, or partially included multi-modality image registration \citep{journal/EIJ/el2016,journal/CRP/khalil2018}.
Cardiac image fusion survey works mainly focused on a specific CVD \citep{journal/JBR/piccinelli2013} or intra-modality fusion \citep{journal/IPADCD/chauhan2021}.
Though there are several cardiac segmentation review works \citep{journal/MedIA/petitjean2011,journal/JHE/zhuang2013,journal/MAGMA/peng2016,journal/FCM/chen2020}, they only concentrated on single-modality or single-structure cardiac imaging.
In contrast, we provide a comprehensive review of  multi-modality cardiac image analysis, excluding single-modality works. 

\subsection{Structure of this review} \label{intro:review structure}
The rest of the survey is organized as follows.
Section \ref{cardiac imaging} summarizes widely used imaging modalities for cardiology.
Section \ref{cardiac computing} covers the well-developed computing algorithms.
In Section \ref{cardiac data and evaluation}, publicly available datasets and performance evaluation measurements are listed. 
Discussion of current clinical applications, challenges and future perspectives of multi-modality cardiac image computing is given in Section \ref{discussion}, followed by a conclusion in Section \ref{conclusion}.

\section{Cardiac imaging} \label{cardiac imaging}

Many imaging techniques are available for cardiac analysis, as presented in \Leireffig{fig:intro:MM-cardiac images}.
In general, these techniques either focus on the anatomical (wall thickness, volume, coronary artery morphology, etc.) or the functional (wall motion patterns, cardiac perfusion, etc.) aspects.
This section briefly reviews the most commonly used imaging techniques for clinical investigation of the heart.

\subsection{X-ray (fluoroscopy) } \label{imaging:x-ray} 
X-ray imaging captures a projection of a 3D structure on a 2D plane, with poor soft-tissue contrast but providing high contrast images of devices \citep{conf/MICCAI/housden2012}.
It has an important role in the preliminary assessment of CVDs.
For example, increased heart size and the presence of increased pulmonary vascular markings or pleural effusions in X-ray may suggest pulmonary congestion secondary to heart failure.
X-ray angiography can provide information about coronary arteries (CA) and their stenosis, and therefore is a standard and popular assessment tool for medical diagnosis of coronary artery diseases (CADs) \citep{journal/JCS/tayebi2015}. 
It is however fundamentally limited by its 2D representation, as extensive 3D/ 4D information about the CA is lost.
Though many 3D reconstruction methods have been proposed for CA reconstruction from 2D X-ray angiography, it remains a challenging and dynamic research area \citep{journal/MedIA/ccimen2016}.

X-ray fluoroscopy is a real-time imaging technique that displays a continuous X-ray image on a monitor and is easy to use during interventions \citep{journal/CiC/ma2010,journal/MP/faranesh2013}.
Therefore, it has been routinely utilized in patients to guide vascular and cardiac interventions.
For example, it can visualize the blood flow through the CA for arterial occlusion detection \citep{journal/MedIA/ma2020}, guide placement of pacemaker leads for cardiac resynchronization therapy (CRT) \citep{journal/CiC/ma2010} and enhance image guidance during cardiovascular interventional procedures \citep{journal/MP/faranesh2013}.
However, it is limited by exposure to ionizing radiation, poor soft tissue characterization and lack of quantitative functional information.

\begin{figure*}[t]\center
 \includegraphics[width=0.99\textwidth]{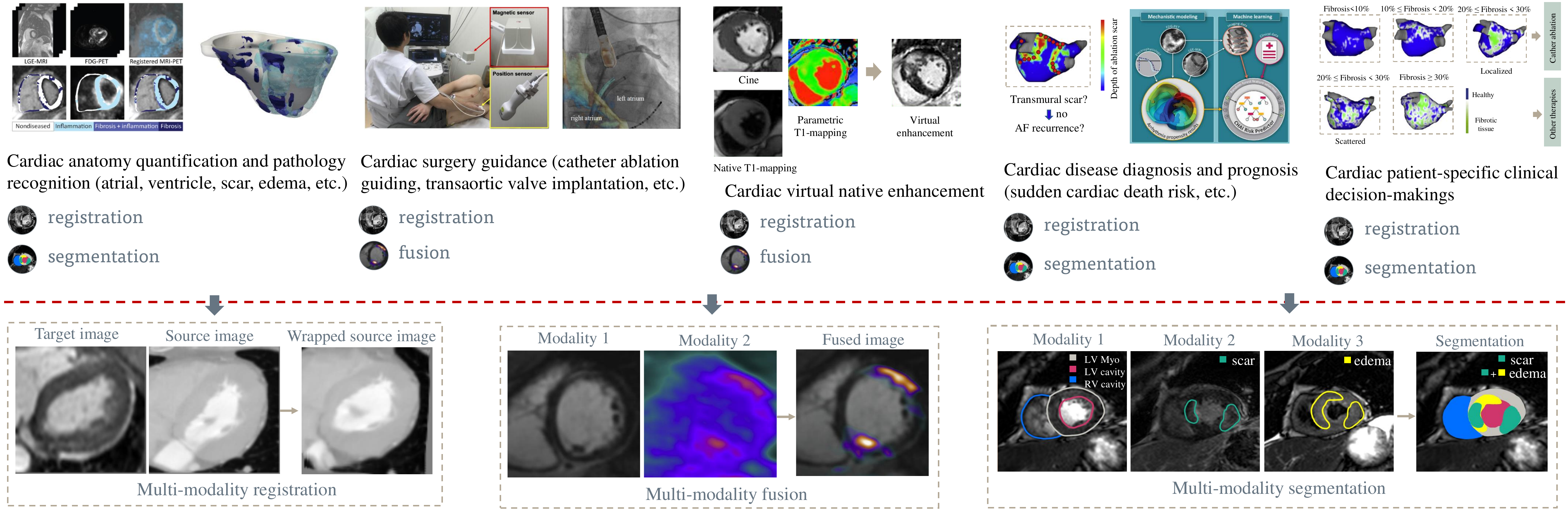}\\[-2ex]
  \caption{Multi-modality cardiac image computing tasks motivated by corresponding clinical applications.
  Here, the images in the clinical application are adapted from \citet{journal/SA/shade2021}, \citet{conf/MICCAI/housden2012}, \citet{journal/Cirir/zhang2021} and \citet{journal/JACC/siebermair2017} with permission;
  and the images in the multi-modality fusion and segmentation are adapted from \citet{journal/HFR/quail2017} and \citet{journal/MedIA/li2022-myops} with permission, respectively.
  }
\label{fig:method:task}
\end{figure*}

\subsection{Computed tomography} \label{imaging:CT} 
Cardiac CT generates cross-sectional images of the human body, with excellent temporal and spatial resolution.
It represents the X-ray attenuation properties of the tissue being imaged, where the measured attenuation can be employed to reconstruct the 3D heart structure.
Moreover, contrast agents can be utilized in CT to enhance the 3D visualization of CA \citep{journal/European/mylonas2014}, the localization of the myocardial infarction (MyoI) \citep{journal/AJR/bauer2010}, the detection of LA appendage (LAA) thrombus in atrial fibrillation (AF) patients \citep{journal/CCI/spagnolo2021}, and the estimation of myocardial (Myo) blood flow \citep{journal/CCI/alessio2019}.
CT angiography (CTA) can give complementary information (clear anatomical information) about the CA tree but without any functional information in cardiac imaging \citep{journal/CMIG/zhou2012}.

\subsection{Magnetic resonance imaging} \label{imaging:MRI} 
Cardiac MRI is an imaging technique with high resolution as well as high tissue contrast for the assessment of heart chambers, cardiac valves, cardiac vessels, and surrounding structures such as the pericardium \citep{journal/JMRI/earls2002}.
It can also be used to study myocardial perfusion and viability, and permits the calculation of ventricular function parameters such as left ventricle (LV) ejection fraction, end diastolic/ systolic volumes or LV mass.
Compared to nuclear imaging (PET or SPECT), the dependence of MRI signal on regional hypoperfusion is minimal \citep{journal/JEI/kang2012}.
Note that within MRI, various pulse sequences can attenuate different tissue features to identify anatomical and functional information, resulting in images of varying contrasts.
For example, cine MRI can provide clear structure and global ventricular performance via quantification of chamber volumes and function;
late gadolinium enhanced (LGE) MRI provides fibrosis information by visually enhancing fibrosis regions;
and T2-weighted images can be used to identify myocardial edema caused by inflammation or acute ischemia \citep{journal/TPAMI/Zhuang2019}.
In clinical practice, myocardial analysis of LGE/ T2-weighted MRI is typically combined with cine MRI considering the poor image quality of LGE/ T2-weighed MRI \citep{journal/JACC/kim2009}.
Similar to CTA, MR angiography is a standard technique for imaging the aorta and large vessels of CA \citep{journal/AFP/mieres2007}.


\subsection{Echocardiography} \label{imaging:US}
Echo (or US of the heart) is a real-time cardiac imaging technique without irradiation, and its temporal resolution is excellent.
Echo is the most widely used imaging technique in cardiology as it is relatively inexpensive, radiation-free and portable.
Dynamic Echo is the gold standard modality to assess myocardial strain.
The limitations of Echo include small FOV, poor signal-to-noise ratio, artifacts, and high sensitivity to operators and patients.
There are various echocardiographic modalities, such as transthoracic Echo (TTE), transoesophageal Echo (TEE), and intracardiac Echo (ICE) \citep{journal/ACD/hascoet2016}.
TTE can be used to identify the great majority of cardiac causes of shock \citep{journal/Stroke/de2006}, diagnose pulmonary embolism \citep{journal/JASE/fields2017}, and evaluate patients with suspected myocarditis \citep{journal/AJR/goitein2009}. 
TEE allows for high-quality color flow imaging, which can be utilized for visualizing the cardiac structures near the upper chambers and assisting device positioning and deployment during cardiac treatment procedures \citep{journal/CRP/khalil2018}.
However, TEE is an invasive imaging technique, as it is acquired via a thin tube connected with an Echo probe through the mouth of patients into the esophagus.
ICE is a catheter-based technique that presents imaging within the heart as well as aortic root images, with a larger view than TEE.
Several research works have presented the superiority of 3D over 2D Echo alone in the catheterization laboratory \citep{journal/ACD/martin2013,journal/EHJ/hascoet2015}.
Unfortunately, 3D Echo is not widely available in clinical practice and is more expensive compared with 2D Echo.

\subsection{Positron emission tomography} \label{imaging:PET} 
Cardiac PET is a nuclear medical imaging technique that can measure myocardial perfusion, metabolism and regional/ absolute myocardial blood flow \citep{journal/JNC/slomka2009,journal/HFR/angelidis2017}.
It is particularly suitable for obese patients and patients with multi-vessel diseases to confirm or exclude balanced myocardial ischemia \citep{journal/NRC/dewey2020}. 
Its spatial resolution (about 3-4 mm) is lower compared to cardiac MRI or CT (about 1 mm), so PET usually can not provide clear anatomical information about the heart.
Therefore, CT is usually utilized to perform attenuation correction for better resolution of PET images.

\subsection{Single photon emission computed tomography} \label{imaging:SPECT} 
SPECT is the most commonly employed imaging technique for clinical myocardial perfusion, whereas PET is the clinical gold standard for myocardial perfusion quantification.
With new detectors, SPECT can quantify myocardial blood flow and is suitable for patients with a high body mass index (BMI) \citep{journal/NRC/dewey2020}.
SPECT can also be used to detect ischemia and viability in heart failure \citep{journal/HFR/angelidis2017}, assess the physiological relevance of coronary lesions and provide a high prognostic predictive value \citep{journal/SNM/kaufmann2009}. 
Moreover, compared with PET, SPECT can accurately assess myocardial viability and has a lower cost of equipment and less expensive radiotracers, but with lower resolution imaging prone to artifacts and attenuation \citep{journal/HFR/partington2011}.

\section{Multi-modality cardiac image computing} \label{cardiac computing}
With these cardiac imaging techniques, one could combine their advantages for cardiac analysis in the clinic, e.g. for cardiac pathology identification, cardiac surgery guidance, and cardiac disease diagnosis and prognosis, as presented in \Leireffig{fig:method:task}.
To achieve this, multi-modality cardiac image computing is required, mainly including multi-modality registration, fusion and segmentation.
In this section, we summarize techniques for the three multi-modality cardiac image computing tasks.

\subsection{Registration}

Multi-modality image registration is a native way to propagate knowledge between modalities as well as a primary step in integrating information of multi-modality images.
It can align images with respect to each other and typically involves several components, i.e., feature extraction (optional), geometrical transformation, similarity measure, and optimization.
Accurate multi-modality image registration can offer additional crucial clinical relevant information, which may be unavailable in a single imaging modality.
However, registration is an ill-posed  problem with several challenges, such as anatomy variations, possible partial overlaps between images, and high variability of tissue appearance under different modalities that leads to the lack of robust similarity measures.

Cardiac image registration is particularly challenging due to respiration and mixed motions during the cardiac cycles.
Furthermore, multi-modality cardiac images are normally inconsistent in the cardiac phases, slice thickness, dimension, and state (static or dynamic), which introduce additional challenges.
For example, for multi-sequence MRI registration, T2 imaging generally only includes a few slices with larger thickness compared to bSSFP and LGE imaging \citep{journal/TPAMI/Zhuang2019}.
Echo / CT image registration frameworks usually consist of two major steps: temporal synchronization and spatial registration \citep{journal/TMI/huang2009,journal/MBEC/khalil2017}.
Temporal synchronization allows the echocardiographic time-series data to be time-stamped to identify frames that are in a similar cardiac phase to the CT volume.
Spatial registration aims at producing an interpolated cardiac CT image that matches the Echo image, which sometimes involves a challenging 2D-3D registration.
Despite these challenges, there exist several promising solutions for multi-modality image registration.
We have coarsely divided these methods into three categories, namely intensity based registration, structural/ anatomical information based registration, and image synthesis/ disentangle based registration.
\Leireftb{tb:method:registration} provides a summary of the state-of-the-art cardiac multi-modality registration methods according to this classification in chronological order.

\begin{table*} [htbp] \center
    \caption{Summary of previously published works on the \textit{multi-modality cardiac image registration}.
    reg: registration; seg: segmentation;
    MS-CMRSeg: paired cardiac bSSFP, T2-weighed, and LGE MRIs from Multi-sequence Cardiac MR Segmentation Challenge \citep{journal/MedIA/zhuang2019};
    X-rayAI: X-ray angiography image; CTA: CT angiography; CSA:coronary sinus angiogram; EAM: electroanatomical mapping;
    MI: mutual information; NMI: normalized MI; NCC: normalized cross correlation; RMS: root mean squared; MSD: mean square distance; ECC: entropy correlation coefficient; JE: joint entropy; PSMI: point similarity measure based on MI; CoR: correlation ratio; WC: Woods criterion; SEMI: spatially encoded MI; VWMI: Viola-Wells MI; MMI: Mattes MI; SSD: sum of squared difference; WJS: weighted Jensen-Shannon; KL: Kullback-Leibler; LCC: local correlation coefficient; MD: mean distance; ASD: average surface distance;
    CA: coronary arteries; CoC: correlation coefficient;
    FFD: free-form deformation; Syn/ Dis: image synthesis/ disentangle based registration. 
     }
\label{tb:method:registration}
{\small
\scalebox{0.84}
{
\begin{tabular}{p{0.1cm}p{4cm}|p{2.5cm}p{2cm}p{7cm}p{3cm}}
%
\hline
&Study & Modality & Target & Method & Similarity metric \\ 
\hline
\multirow{13}{*}{\rotatebox{90}{Intensity based registration}} &\citet{conf/FIMH/pauna2003}        & MRI, PET         & WH           & intensity-based rigid reg                      & CoR, CoC, MI \\ 
& \citet{conf/MICCAI/guetter2005}    & CT, PET          & Myo          & intensity-based non-rigid reg                  & KL divergence + MI \\
& \citet{conf/CVPR/cremers2006}      & CT, PET          & Myo          & intensity-based non-rigid reg                  & KL divergence + MI \\
& \citet{conf/MICCAI/guetter2007}    & CT, SPECT        & Myo          & intensity-based rigid reg                      & WJS divergence \\
& \citet{journal/TMI/huang2009}      & CT, Echo         & WH           & intensity-based temporal + 2D-3D rigid reg     & MI \\ 
& \citet{conf/BM/papp2009}           & CT, PET, SPECT   & Myo          & intensity-based reg                            & triple NMI \\
& \citet{journal/SWJ/marinelli2012}  & CT, PET          & Myo          & intensity-based rigid reg                      & MI \\
& \citet{conf/CC/sandoval2013}       & CT, Echo         & WH           & intensity-based rigid + elastic reg            & MI, NMI, ECC, JE, PSMI, CoR, WC \\
& \citet{journal/MP/lamare2014}      & CT, PET          & Myo          & intensity-based affine + elastic reg           & NMI \\
& \citet{journal/TNS/turco2016}      & CT, PET          & WH           & intensity-based rigid reg                      & NMI \\
& \citet{journal/MBEC/khalil2017}    & CT, Echo         & aorta        & intensity-based temporal + rigid reg           & NMI \\
& \citet{journal/MedIA/puyol2017}    & MRI, Echo        & Myo          & intensity-based rigid reg + FFD                & N/A \\
\hline \hline 
\multirow{36}{*}{\rotatebox{90}{Structural/ anatomical information based registration}} & \citet{conf/ISBI/zhang2006}        & MRI, Echo        & LV, RV, Myo  & local phase based affine reg                   & MI \\ 
& \citet{conf/ISBI/cordero2012}      & cine, LGE MRI    & Myo          & local entropy based reg                        & NCC  \\
& \citet{conf/MICCAI/oktay2015}      & CT, Echo         & LV, RV, Myo  & PEM-based rigid + non-rigid reg                & NCC, LCC \\
& \citet{journal/MedIA/choi2016}     & MRI, X-ray       & LV           & normalized gradient field based rigid reg      & NCC \\
& \citet{journal/JNM/kolbitsch2017}  & MRI, PET         & Myo          & motion field based affine reg + FFD            & NCC \\
& \citet{journal/CMBBE/gouveia2017}  & CT, X-ray        & CA           & regression based rigid reg                     & N/A \\ 
& \citet{journal/JCARS/atehortua2020}& MRI, Echo        & Myo          & salient image based Eulerian and nonrigid reg  & NCC \\
& \citet{conf/EJNM/savi1995}         & Echo, PET        & Myo          & landmark-based rigid reg                       & MSD \\ 
& \citet{journal/MedIA/makela2003}   & MRI, PET, Echo   & LV, RV, Myo  & landmark and seg based reg                     & region energy \\ 
& \citet{journal/JCI/walimbe2003}    & Echo, SPECT      & Myo          & landmark-based 2D-3D reg                       & MI \\ 
& \citet{journal/Cir/de2006}         & MRI, X-ray       & LV, RV, Myo  & landmark-based rigid reg                       & MSD \\ 
& \citet{journal/MedIA/baka2013}     & CT, X-ray        & CA           & landmark-based rigid reg                       & 2D-3D distance \\ 
& \citet{journal/MP/aksoy2013}       & CT, X-ray        & CA           & landmark-based rigid reg                       & 2D-3D distance \\ 
& \citet{journal/EP/doring2013}      & Echo, CSA        & Myo          & landmark-based reg                             & N/A \\
& \citet{conf/MIC/smith2014}         & PET, SPECT       & Myo          & landmark-based reg                             & N/A \\
& \citet{journal/JCS/tayebi2015}     & CT, X-rayAI      & CA           & landmark-based affine reg                      & RMS \\
& \citet{journal/JMI/khalil2017}     & CT, Echo         & LA           & landmark-based rigid reg                       & NMI \\
& \citet{journal/Radio/faber1991}    & MRI, SPECT       & LV           & point set based rigid reg                      & MD \\ 
& \citet{journal/AJR/sinha1995}      & MRI, PET         & Myo          & point set based rigid reg                      & MSD \\ 
& \citet{conf/MICCAI/santarelli2001} & MRI, SPECT       & Myo          & point set based non-rigid reg                  & MSD \\
& \citet{journal/JCI/sturm2003}      & MRI, CTA         & Myo          & point set based rigid reg                      & MSD \\
& \citet{journal/TMI/tavard2014}     & CT, Echo, EAM    & LV           & point set based temporal reg                   & SSD \\
& \citet{journal/JCARS/peoples2019}  & CT, Echo         & WH           & point set based non-rigid reg                  & posterior \\
& \citet{journal/CMBBE/scott2021}    & CT, Echo         & WH           & point set based rigid reg                      & N/A \\
& \citet{conf/MICCAI/makela2001}     & MRI, PET         & Myo          & seg based rigid reg                            & N/A \\
& \citet{journal/MP/woo2009}         & CT, SPECT        & Myo          & seg based rigid + nonrigid reg                 & SSD \\
& \citet{conf/ISBI/courtial2019}     & cine MRI, CT     & WH           & seg based rigid reg                            & NMI \\
& \citet{journal/TPAMI/Zhuang2019}   & MS-CMRSeg        & Myo          & seg-based rigid reg + FFD                      & posterior \\ 
& \citet{journal/MedIA/li2020}       & bSSFP, LGE MRI   & WH           & seg-based affine reg + FFD                     & SEMI \\
& \citet{conf/MICCAI/ding2020}       & MRI, CT          & Myo          & seg based non-rigid reg                        & Dice \\
& \citet{conf/MICCAI/luo2020}        & MS-CMRSeg        & LV, RV, Myo  & seg based rigid + non-rigid reg                & Dice \\ 
& \citet{conf/STACOM/ding2021}       & MRI, CT          & Myo          & seg based non-rigid reg                        & SEGI \\ 
& \citet{journal/MRM/guo2021}        & cine, LGE MRI    & Myo          & seg based affine + non-rigid reg               & N/A \\ 
& \citet{journal/CMIG/gilardi1998}   & PET, SPECT       & Myo          & surface based rigid reg                        & ASD \\ 
& \citet{journal/JNM/martinez2007}   & CT, PET          & Myo          & outline based spatial reg                      & NMI \\
& \citet{journal/MedIA/betancur2016} & cine/ LGE MRI, Echo& Myo        & outline based rigid reg                        & MSD, NCC, VWMI, MMI, NMI\\
\hline \hline 
\multirow{4}{*}{\rotatebox{90}{{Syn/ Dis}}} & \citet{journal/TMI/shi2012}        & cine, tagged MRI & Myo          & pseudo-anatomical MRI based nonrigid reg       & weighed NCC \\ 
& \citet{conf/MIC/dey2012}           & CT, SPECT        & WH           & synthetic CT based rigid and non-rigid reg     & SSD \\
& \citet{conf/IPCAI/li2013}          & CT, Echo         & LV           & synthetic CT based non-rigid reg               & MI \\ 
& \citet{journal/TMI/chartsias2020}  & bSSFP/ LGE MRI   & LV, Myo      & disentangled representation based affine reg   & MI \\  
\hline 
\end{tabular}}} \\
\end{table*}

\subsubsection{Intensity based registration}
A straightforward approach is to employ modality invariant similarity metrics, such as mutual information (MI) and its variants such as normalized MI \citep{journal/TMI/pluim2003,conf/CC/sandoval2013}.
MI and its variations are well suited for multi-modality image registration, as they do not assume a linear relationship between intensities of different imaging modalities \citep{journal/TMI/maes1997}.
Kullback–Leibler (KL) divergence has also been employed as a similarity metric for multi-modality cardiac image registration \citep{conf/CVPR/cremers2006,conf/MICCAI/guetter2005}.
It enforces the joint intensity distribution of the source and target images according to the priors of learned distributions from the pre-aligned images.
The prior knowledge can be obtained from the expert knowledge of a physician who manually aligns the images.
Also, one can leverage the fused imaging data acquired using dual-modality/ hybrid scanners, which can offer extensive amounts of pre-registered data \citep{conf/MICCAI/guetter2005}.
Jensen-Shannon divergence provides a more suitable measure than KL divergence in quantifying histogram discrepancy \citep{conf/MIVR/liao2006}. 
In addition, it is upper bounded and symmetric, facilitating its use for multi-modality image registration, and its weighted version can further ensure an organ-specific intensity co-occurrence \citep{conf/MICCAI/guetter2007}.

However, such global intensity-based similarity measures often ignore the spatial information of the target, so they are unsuitable in some situations.
For instance, the tissue presented in  Echo is inherently characterized by a specific spatial distribution of speckles rather than a specific distribution of gray scales that exist in MRI and CT \citep{conf/CC/sandoval2013}.

\begin{figure}[t]\center
 \includegraphics[width=0.48\textwidth]{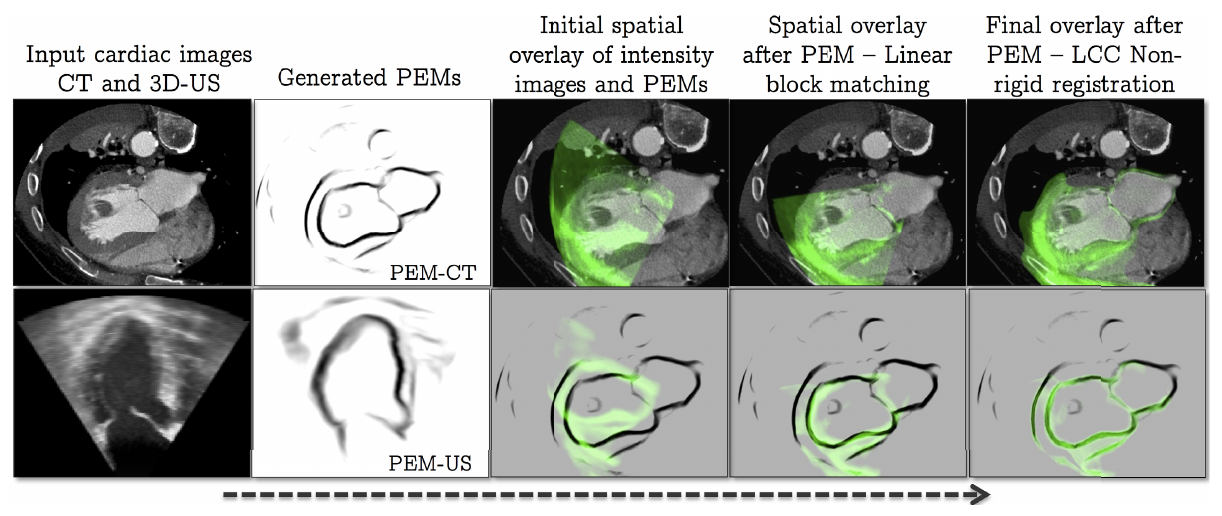}\\[-2ex]
   \caption{An example of multi-modality cardiac image registration. Here, a generic and modality-independent image representation, namely probabilistic edge map (PEM), was proposed for cardiac CT and US registration. 
   The PEM, by only highlighting the registration of relevant image structures, i.e., LV and atrial boundaries, is a more robust approach than registering every voxel in the images.
   Image adapted from \citet{conf/MICCAI/oktay2015} with permission.}
\label{fig:method:mm-reg}
\end{figure}

\subsubsection{Structural/ anatomical information based registration} 
Instead of directly employing image intensities, one could utilize structural information, which is supposed to be independent to modalities.
In this way, the influence of appearance differences among various modalities can be mitigated, and a simple L1 or L2 distance can be used for the multi-modality image registration.
One could extract image representations, namely structural representations, such as entropy, Laplacian images \citep{journal/MedIA/wachinger2012}, local phase images \citep{conf/ISBI/zhang2006,journal/TMI/grau2007}, and probabilistic edge maps (PEMs) \citep{conf/MICCAI/oktay2015}.
Entropy images satisfy only certain requirements of a relaxed version of the theoretical properties, but they are computationally fast and lead to good alignment, facilitating its use as a practical solution.
Laplacian images capture the intrinsic structure of imaging modalities, with the preservation of optimal locality.
The local phase image can be a characterization of features detected in a signal, such as edges and ridges in a signal, so it has been shown appropriately for registration involving Echo images.
In contrast to local phase representations, PEMs only highlight the image structures that are relevant for image registration, as presented in \Leireffig{fig:method:mm-reg}.
Note that several anatomical structures that are visible in US images may not be visible in CT/ MR images and vice versa, such as endocardial trabeculae \citep{conf/MICCAI/oktay2015}.
Therefore, only registering the region of interest (ROI) can be more robust than registering every voxel in the images.

Modality-invariant anatomical landmarks, surfaces (contours or point clouds) or labels can also be used to define the transformation from one image to the other due to their inherent correspondences with original images.
For example, \citet{conf/EJNM/savi1995,journal/EP/doring2013,conf/MIC/smith2014} employed the landmarks being identified in multi-modality images for registration.
\citet{conf/MICCAI/ding2020} predicted a dense displacement field (DDF) for cardiac CT and MRI registration by aligning the anatomical labels of MRI and CT instead.
Nevertheless, these methods are highly sensitive to the accuracy of segmentation or landmark, so manual segmentation or landmark detection is usually required, which is impractical in image-guided intervention.

\subsubsection{Image synthesis/ disentangle based registration}
An alternative way is to reduce the multi-modality registration into a mono-modality problem, where most existing mono-modality registration methods then can be applied \citep{conf/IPCAI/li2013}.
One could either synthesize one modality from another one or transfer multiple modalities into a common domain \citep{journal/MedIA/chen2017,conf/CVPR/arar2020}.
The main idea behind this is to mitigate the large appearance gap among different modalities.
To achieve this, we can take advantage of prior knowledge of the physical properties of imaging devices \citep{journal/TMI/roche2001} or capture their intensity relationships \citep{journal/MedIA/cao2017}.
As for mapping multiple modalities to a common space, these modalities need to share the same anatomical structure or features in order to build a meaningful correspondence.
A major issue of current image synthesis based registration methods is that the image synthesis is normally performed in a single direction.
Specifically, current image synthesis is often adapted from the imaging modality with rich anatomical details (e.g., MRI) to the imaging modality with limited anatomical details (e.g., CT/ US).
Note that it is quite challenging to achieve accurate local deformation prediction based on an imaging modality with limited anatomical details, especially in the presence of large local deformations in cardiac images.
One could also disentangle the anatomy and modality information from images, and then a mono-modality registration can be adopted on the latent embedding space \citep{journal/TMI/chartsias2020}.
A major issue with current disentanglement based methods is that they generally can not explicitly impose the disentanglement, so the learned representations may be exposed to information leakage \citep{conf/ipmi/ouyang2021}.

\subsubsection{Discussion}
Multi-modality cardiac image registration is a crucial step for further cardiac analysis, providing complementary information for cardiac image fusion (Sec \ref{method:fusion}) and facilitating the cardiac image segmentation (Sec \ref{method:segmentation}).
It allows for a more comprehensive analysis of the cardiac functions and pathologies, and has many applications in the clinic, such as monitoring the progression of diseases, quantifying the effectiveness of treatment mechanisms, surgery planning, and intra-operative navigation \citet{journal/EP/doring2013,conf/NIMPR/giannoglou2006,journal/MBEC/khalil2017}.
During the registration procedure, different combinations of dimensions, such as 2D-to-2D, 3D-to-3D, 2D-to-3D, and 3D-to-4D, may be involved, and they usually have different applications and optimization targets.
For example, 2D-to-3D image registration is normally employed for establishing correspondence between X-ray and 3D volumes.
Its main application is to facilitate image-guided surgery \citep{conf/ICASI/khalil2017,journal/Cir/dori2011}, so computational complexity must be minimized without compromising accuracy.

\begin{figure}[t]\center
 \includegraphics[width=0.49\textwidth]{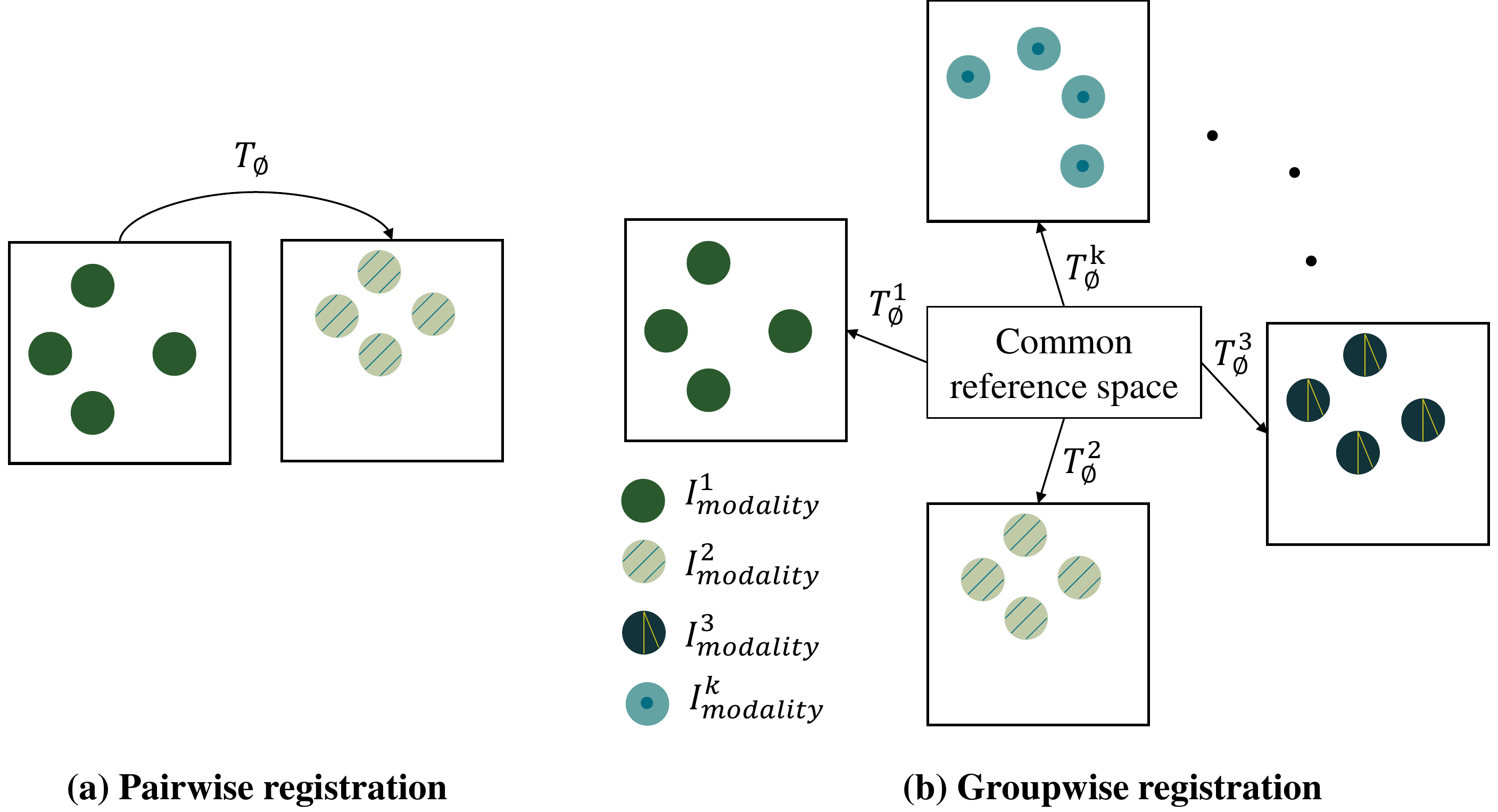}\\[-2ex]
   \caption{Graphical illustration for (a) pairwise registration (b) groupwise registration.
   Here, $T_{\phi}$ is the adopted transformation, and $I_{modality}^{1, 2, 3, ..., k}$ refer to different imaging modalities.
   Here, each square box represents an image, the distribution of the circles in the box indicates the image anatomy, and the different textures of the circles represents different imaging modalities.
   }
\label{fig:discussion:groupwise}
\end{figure}

We have summarized various cardiac image registration methods according to the key elements of methodologies.
One can see that structural/ anatomical information based registration is the most popular for multi-modality cardiac image registration.
This can be attributed to the fact that the structural/ anatomical information among different modalities is the same, ensuring the accuracy of registration.
Also, structural/ anatomical information based registration can use simple similarity metrics instead of computationally expensive MI-based metrics, and thus accelerate the registration procedure.
Pure rigid transformations have been applied in several works, but rigid cardiac image registration is generally not enough to describe the spatial relationship due to the existence of cardiac motion.
Therefore, there is considerable research extending the transformation to the combination of rigid and non-rigid ones.

Conventional cardiac image registration basically is an iterative-based optimization procedure, which could be quite slow, especially for non-rigid image registration.
DL-based registration procedures can be computationally efficient \citep{journal/CEE/boveiri2020}, and have been applied in multi-modality cardiac images \citep{journal/CMBBE/gouveia2017,conf/MICCAI/ding2020,conf/MICCAI/luo2020}.
These methods generally require extensive anatomical labels for supervised network training.
To solve this, \citet{conf/STACOM/ding2021} investigated unsupervised DL-based cardiac multi-modality image registration by introducing a modality-invariant structural representation, i.e., spatially encoded gradient information.
One can also exploit the image similarity analogous to conventional intensity-based image registration for unsupervised DL-based cardiac image registration \citet{journal/MedIA/de2019};
or employ image disentangle learning to embed an image onto a domain-invariant latent space for the registration \citep{conf/IPMI/qin2019}.  
Also, groupwise registration has recently emerged and has been applied for multi-modality registration and segmentation of cardiac images \citep{journal/TPAMI/kouw2019,journal/TPAMI/Zhuang2019,conf/MICCAI/luo2020}.
Compared to pairwise registration, groupwise registration is able to handle several imaging modalities simultaneously in an unbiased way (see \Leireffig{fig:discussion:groupwise}).
Therefore, DL-based multi-modality image registration can be further facilitated by introducing groupwise learning.
Nevertheless, there are no general automatic methods due to the wide variety of modalities and clinical scenarios in cardiology.

\subsection{Fusion} \label{method:fusion}

\begin{figure}[t]\center
 \includegraphics[width=0.4\textwidth]{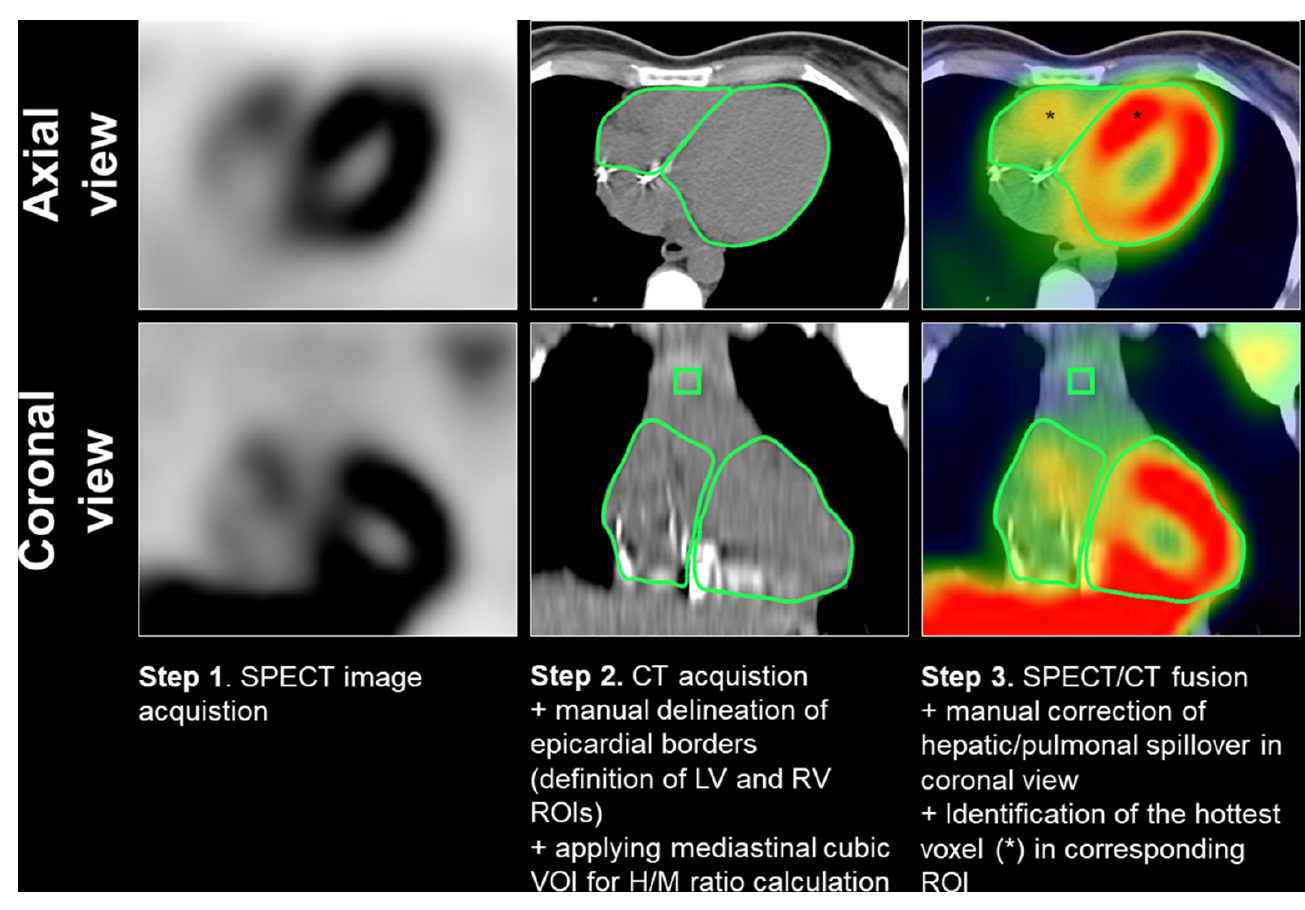}\\[-2ex]
   \caption{An example of multi-modality cardiac image fusion. Here, cardiac CT and SPECT are fused to assess inherited arrhythmia syndromes. Image adapted from \citet{journal/IJC/siebermair2020} with permission.}
\label{fig:method:mm-fusion}
\end{figure}

\begin{figure}[t]\center
 \includegraphics[width=0.46\textwidth]{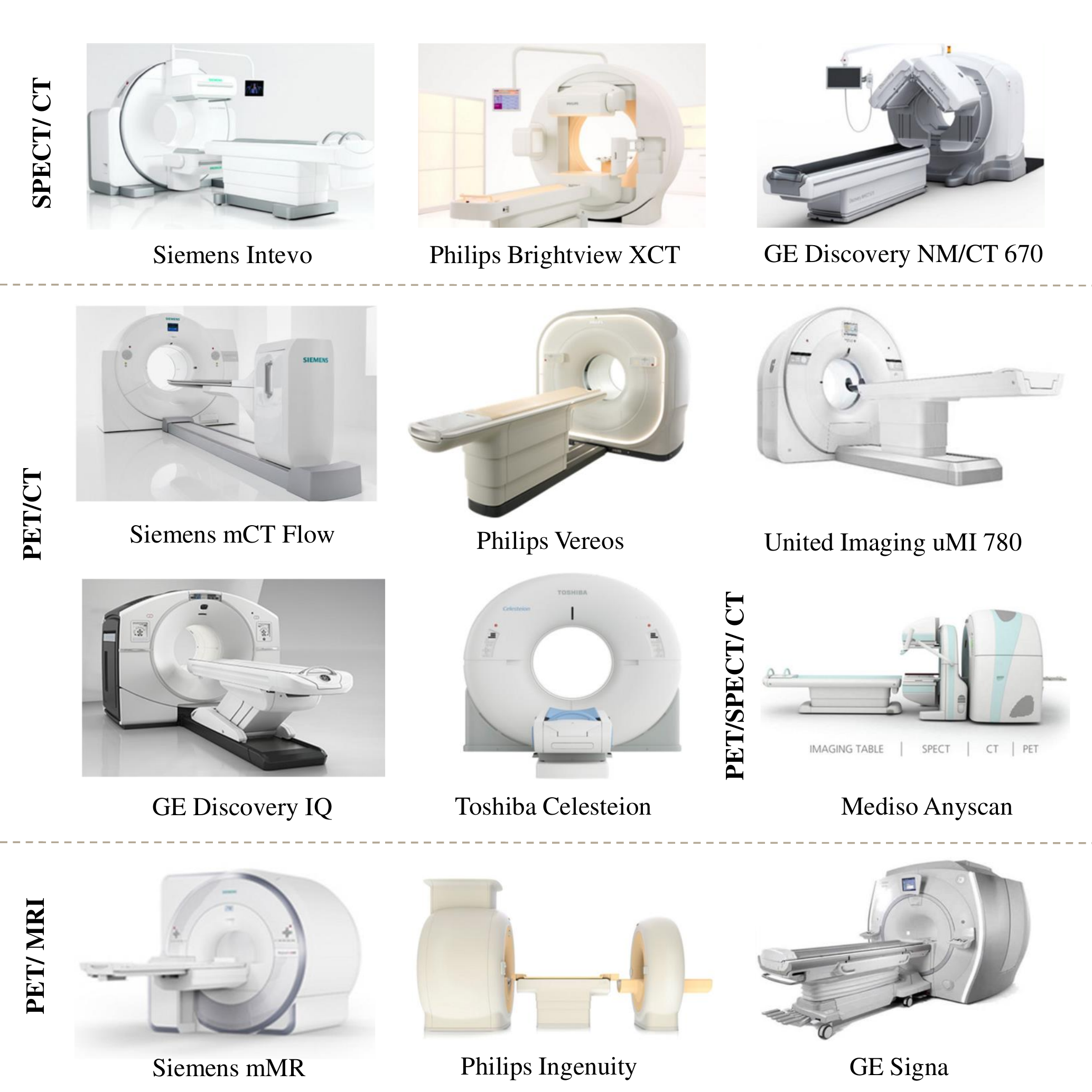}\\[-2ex]
  \caption{Examples of commercially available hybrid imaging devices for clinical use. Image modified from \citet{journal/FiP/cal2018} with permission.}
\label{fig:discussion:hybrid device}
\end{figure}

Multi-modality cardiac image fusion aims at integrating relevant information from several images into a single one, to obtain a more informative and representative integrated image, as presented in \Leireffig{fig:method:mm-fusion}.
The fusion process can improve image quality and reduce randomness and redundancy, in order to augment clinical accuracy and robustness of the analysis for an accurate diagnosis.
For decades, images from different modalities have been used to develop image fusion frameworks either invasive or non-invasive, anatomical and functional, leading to the emergence of hybrid devices such as PET/ MRI, PET/ CT and SPECT/ CT \citep{journal/JNC/piccinelli2020}, as shown in \Leireffig{fig:discussion:hybrid device}.
Specifically in the field of cardiology, image fusion is mainly used to integrate functional imaging, such as SPECT or PET, and anatomical imaging, such as CT or MR.
Also, it permits a combination of the advantages of each modality, such as excellent soft-tissue contrast and/ or higher spatial resolution (CT or MRI) and high temporal resolution (X-ray fluoroscopy, Echo) for detailed real-time feedback.
Among these hybrid cardiac imaging techniques, PET (or SPECT)/ CT (CTA) is currently the most widespread \citep{journal/CGF/lawonn2018}.
As cardiac MRI and CT are both high-resolution images, their fusion is relatively rare and could be more challenging \citep{journal/CGF/lawonn2018}.
\Leireftb{tb:method:fusion} summarizes the applied modalities in cardiac image fusion with their applications in cardiac imaging studies.
Note that as some of these reviewed papers provide limited details about fusion methodology, we classified image fusion works based on their applied modalities instead of the methodologies.
Also, we only include parts of representative image fusion studies without pretending to be exhaustive.

\begin{table*} [htbp] \center
    \caption{Summary of previously published works on the \textit{multi-modality cardiac image fusion}.
     PV: pulmonary vessel; PC: pulmonary conduit; PA: pulmonary artery; LSVC: left superior vena cava; MV: mitral valve; MyoI: myocardial infarction;
    CCAA: complex CA anomalies; PCI: percutaneous coronary intervention; AF: atrial fibrillation; TAVI: trans-aortic valve implantation;
    CHD: congenital heart disease; CRT: cardiac resynchronization therapy; MBF: myocardial blood flow; MFR: myocardial flow reserve; RFR: relative flow reserve;  iAS: inherited arrhythmia syndromes; SHD: structural heart disease; EM: electromagnetic; STDME: simultaneous two-screen display of multidetector-CT and real-time echogram; TE: transendocardial; VNE: virtual native enhancement; MSF: multi-sequence MRI fusion. 
     }
\label{tb:method:fusion}
{\small
\scalebox{0.88}
{
%
\begin{tabular}{p{0.1cm}p{4cm}|p{2.5cm}p{1.6cm}p{4cm}p{6cm}}
\hline
& Study & Modality & Target & Method & Motivation/ Application \\ 
\hline
\multirow{15}{*}{\rotatebox{90}{PET/ SPECT and CT/ MR fusion}} & \citet{journal/CHD/grani2017}        & PET, CT        & CA          & CardIQ fusion software     & assess the impact of CCAA on Myo perfusion \\
& \citet{journal/RSM/aguade2017}       & PET, CT        & Aorta       & N/A                        & diagnose and evaluate the extent of endocarditis \\
& \citet{journal/JNC/piccinelli2020a}  & PET, CTA       & Myo         & rigid reg                  & extract MBF, MFR and RFR along CAs \\
& \citet{journal/HFR/quail2017}        & PET, MRI       & Myo         & N/A                        & visualize each CA and strain in its territory \\
& \citet{journal/medR/izquierdo2020}   & PET, MRI       & Myo         & OsiriX DICOM viewer        & explore the effect of histone deacetylases on the heart \\
& \citet{journal/JNC/zandieh2018}      & PET, MRI       & Myo         & landmark based rigid reg   & diagnose patients with cardiac sarcoidosis \\
& \citet{journal/CJ/degrauwe2017}      & PET, MRI, CT   & Myo, PV     & N/A                        & assess cardiac paragangliomas \\
& \citet{journal/EJNMMI/gaemperli2007} & SPECT, CT      & Myo         & CardIQ fusion software     & combine information of CA and lesion \\
& \citet{journal/IJR/sazonova2017}     & SPECT, CT      & PC          & hybrid devices             & diagnose infectious endocarditis \\
& \citet{journal/IJC/siebermair2020}   & SPECT, CT      & LV, RV      & N/A                        & evaluate chamber-specific patterns of autonomic innervation in iAS \\
& \citet{journal/EJR/koukouraki2013}   & SPECT, CTA     & Myo         & CardIQ Fusion software     & clinical management of patients with suspected CAD \\
& \citet{journal/JACC/nakahara2016}    & SPECT, CTA     & Myo         & fusion based bull’s eye    & determine hemodynamically relevant coronary vessels \\
& \citet{journal/JNM/kiricsli2014}     & SPECT, CTA     & Myo         & point-cloud based reg      & allow Myo perfusion defect correlations with corresponding CA \\
& \citet{journal/JNC/piccinelli2018}   & SPECT, CTA     & Myo         & rigid reg                  & detect and localize CAD \\
& \citet{journal/EJHI/yoneyama2019}    & SPECT, CTA     & Myo         & manual registration        & defect correlations with corresponding CA \\
\hline \hline
\multirow{8}{*}{\rotatebox{90}{Echo and CT/ MRI}} & \citet{journal/EHJCI/maffessanti2017}& Echo, CT       & CA          & N/A                        & visualize each CA and strain in its territory \\
& \citet{journal/TMI/tavard2014}       & Echo, CT, EAM  & LV          & N/A                        & extraction of local electro-mechanical delays \\
& \citet{conf/EMBC/bruge2015}          & Echo, CT, EAM  & LV          & N/A                        & select the LV pacing sites \\
& \citet{journal/JOC/watanabe2021}     & Echo, CT       & CA          & STDME technique            & assess conduit stenosis in complex adult CHD \\
& \citet{conf/IUS/kiss2011}            & Echo, MRI      & LV, RV, Myo & landmark based reg         & assess Myo viability and diagnose ischemia \\
& \citet{conf/IUS/kiss2013}            & Echo, MRI      & LV          & landmark based rigid reg   & guide the echocardiographic acquisition \\
& \citet{journal/EHJ/gomez2020}        & Echo, MRI      & WH          & landmark based reg         & 3D printed heart models \\
& \citet{journal/CMIG/hatt2013}        & Echo, MRI, X-ray & LV        & N/A                        & precise targeting for TE therapeutic delivery \\ 
\hline \hline
\multirow{18}{*}{\rotatebox{90}{X-ray and CT/ MRI/ Echo fusion}} & \citet{journal/JACC/zhou2014}        & X-ray, SPECT   & LV          & vessel-surface rigid reg   & guide LV lead implantation for CRT \\ 
& \citet{conf/STACOM/ma2010}           & X-ray, (LGE) MRI& LV, scar   & manual reg                 & guide LV lead implantation for CRT \\
& \citet{journal/Cir/dori2011}         & X-ray, MRI     & PA          & landmark based rigid reg   & facilitate cardiac catheterization of CHD \\
& \citet{journal/MP/faranesh2013}      & X-ray, MRI     & AO, CA      & affine motion model        & facilitate cardiac interventions \\
& \citet{journal/CCI/abu2014}          & X-ray, MRI     & PA, LSVC    & landmark based reg         & facilitate cardiac catheterization of CHD \\
& \citet{journal/RT/mcguirt2016}       & X-ray, MRI     & PA, LSVC    & landmark based reg         & facilitate cardiac catheterization of CHD \\
& \citet{journal/MedIA/choi2016}       & X-ray, MRI     & LV          & edge based rigid reg       & guide LV lead implantation for CRT \\
& \citet{journal/CCI/grant2019}        & X-ray, MRI     & AO          & landmark based rigid reg   & select transcatheter CHD interventions \\
& \citet{journal/CiC/ma2010}           & X-ray, MRI     & LV          & 2D-3D manual reg           & guide LV lead implantation for CRT \\ 
& \citet{journal/CCI/tomkowiak2011}    & X-ray, (LGE) MRI & LV, scar  & custom software            & guide catheter-based TE delivery \\ 
& \citet{journal/ER/ghoshhajra2017}    & X-ray, CT      & CA          & reg with manual correction & guide chronic total occlusion of PCI \\ 
& \citet{journal/CRC/vernikouskaya2018}& X-ray, CTA     & AO          & reg with manual correction & facilitate TAVI procedures \\ 
& \citet{journal/PMB/ma2010}           & X-ray, Echo    & LV, RV      & intensity-based rigid reg  & guide cardiac catheterization procedures \\
& \citet{journal/CI/clegg2015}         & X-ray, Echo    & MV          & probe tracking             & guide percutaneous SHD interventions \\
& \citet{conf/MICCAI/housden2012}      & X-ray, Echo    & LA, RA      & probe based 2D-3D reg      & facilitate catheter ablation and TAVI \\
& \citet{journal/AIC/zorinas2017}      & X-ray, Echo    & AO          & EchoNavigator system       & assist catheter-based cardiac operations \\
& \citet{journal/ACD/hadeed2018}       & X-ray, Echo    & N/A         & EchoNavigator system       & guide interventional procedures for CHD \\ 
& \citet{journal/CDT/ebelt2020}        & X-ray, Echo    & LAA         & TrueFusion technology      & improve the procedure of LAA closure \\ 
\hline \hline
\multirow{2}{*}{\rotatebox{90}{MSF}} & \citet{conf/ISBI/cordero2012}        & cine, LGE MRI  & Myo         & local entropy based reg    & assess MyoI  \\
 & \citet{journal/Cirir/zhang2021}      & cine, T1 mapping MRI & Myo   & DL VNE generator           & generate gadolinium-free LGE-like MRI \\ 
\hline 
\end{tabular}}
}
\end{table*}

\subsubsection{PET/ SPECT and CT/ MRI fusion} \label{method:fusion:SPECT/PET-CT/MRI}
Healthy cardiac function depends on a close interplay between anatomy and physiology, which is disrupted in many CVDs; thus, their joing evaluation can significantly boost a physician's ability to diagnose and plan treatment \citep{journal/JNC/piccinelli2020}.
Initial attempts to fuse anatomical and functional information date back to the early 1990s and have gradually become established techniques, demonstrating the greater diagnostic power of fused images compared to single-mode or side-by-side interpretation \citep{journal/TBME/peifer1990,journal/JNC/piccinelli2018}.
SPECT and PET can evaluate myocardial perfusion, metabolism, and regional absolute myocardial blood flow, while CT/ MRI provides coronary anatomy and allows for multi-parametric evaluation of cardiac morphology, ventricular function, myocardial perfusion, and viability.
Integrated SPECT/ PET and CT/ MRI allow true simultaneous analysis of the structure and function of the heart and has been applied in clinical practice, as presented in the first row of \Leireftb{tb:method:fusion}.
Compared to PET-CT, hybrid PET-MRI offers some advantages in that it does not require exposure to ionizing radiation from CT and iodinated contrast agents \citep{journal/CCR/bergquist2017}.
It can be applied for the assessment of MyoI \citep{journal/JACC/lee2012}, cardiac sarcoidosis \citep{journal/JNC/zandieh2018}, atherosclerosis \citep{journal/ERMM/cuadrado2016}, non-ischemic cardiomyopathies \citep{journal/EJNMMI/zhang2022}, myocarditis \citep{journal/JNC/nensa2018}, vasculitis \citep{journal/SR/laurent2019}, and cardiac tumors \citep{journal/IJC/rinuncini2016}.
The fusion of SPECT with CT (especially CTA) images is regarded as the most successful example of fusion imaging \citep{journal/JACC/kramer2010}, and has been widely employed for the assessment of CAD \citep{journal/EJR/koukouraki2013,journal/JNC/piccinelli2018}.

\subsubsection{Echo and CT/ MRI fusion} \label{method:fusion:Echo-MRI/CT}
Echo has great potential and offers clear benefits when complementing other modalities due to its ability to provide real-time information.
The limited FOV and noise in Echo data could be alleviated by integrating it with additional imaging modalities.
For example, TEE can be used for the assessment of morphology, function and hemodynamics in most adult congenital heart disease (CHD) patients, while CT enables extensive anatomical analysis due to its high spatial resolution.
Several methods have been developed for synchronized display of real-time Echo images and multi-planar reconstruction images of CT or MRI \citep{journal/JOC/watanabe2021}.
The fused images can provide additional findings, allowing accurate assessment of the diagnosis and severity of diseases, compared to Echo alone. 
For example, it can be employed in cardiac surgery, longitudinal studies, and to derive a more complete picture of the heart (motion from Echo and tissue perfusion from MRI) \citep{journal/JOE/takaya2020}.

\subsubsection{X-ray and CT/ MRI/ Echo fusion} \label{method:fusion:X-ray-MRI}
\begin{figure}[t]\center
    \includegraphics[width=0.45\textwidth]{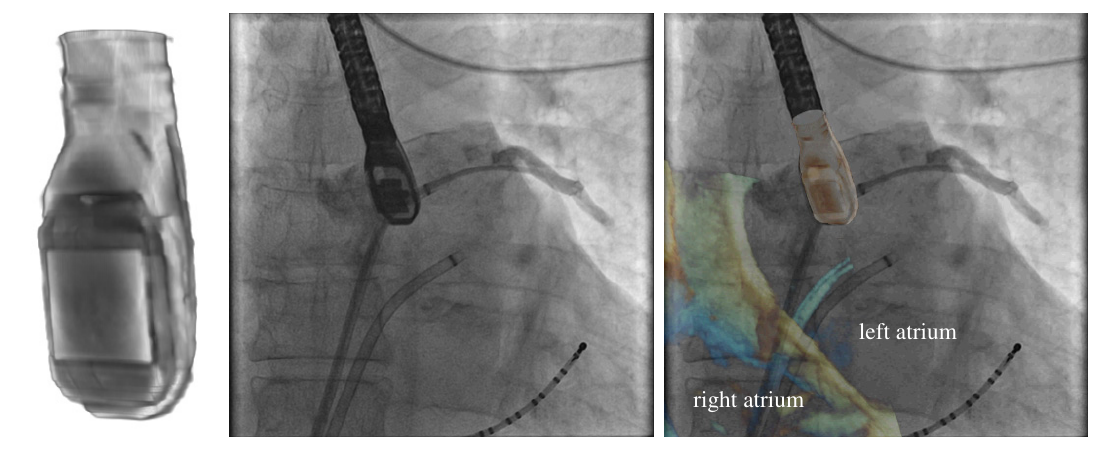}
   \caption{A clinical application example of multi-modality cardiac image fusion for guiding AF catheter ablation and TAVI procedures. 
   Here, X-ray image can present transesophageal Echo (TEE) probe, while TEE is better for visualizing cardiac anatomy. 
   Image adapted from \citet{conf/MICCAI/housden2012} with permission.
   }
\label{fig:method:mm-fusion application}
\end{figure}

X-ray fluoroscopy is commonly used for guiding minimally invasive cardiac interventions due to its excellent catheter and device visualization.
However, it is limited by the 2D projection nature of the images, exposure to radiation, and poor soft-tissue contrast, which however can be solved with the use of diagnostic quality 3D images, such as MRI, CT and real-time 3D Echo.
For example, the fusion of X-ray and MRI can be employed to facilitate cardiac catheterization of CHD patients with significant reductions in contrast and radiation exposure and decide the optimal pacing of the LV lead during CRT procedures.
The introduction of additional LGE MRI during this fusion can further assist cardiologists to avoid scarring regions, as positioning an LV lead within the scarring areas may reduce response to CRT \citep{conf/STACOM/ma2010}.
The fusion of X-ray and CT (or CTA) has been used to guide percutaneous procedures \citep{journal/Cir/dori2011}, guide chronic total occlusion of percutaneous coronary intervention \citep{journal/ER/ghoshhajra2017}, and facilitate trans-aortic valve implantation (TAVI) procedures \citep{journal/CRC/vernikouskaya2018}.  
Real-time 3D stereo echocardiography (such as 3D TEE) can be fused with X-ray fluoroscopy to guide percutaneous structural heart disease interventions \citep{journal/CI/clegg2015} and facilitate AF catheter ablation as well as TAVI procedures \citep{conf/MICCAI/housden2012}. 
As opposed to fluoroscopy, 3D TEE provides excellent detail of 3D anatomy and soft tissue structures and offers ``real-time" intraoperative guidance \citep{journal/CI/clegg2015}.
However, its advantages are limited by Echo shadowing, which reduces visualization of the catheters and metallic structures \citet{journal/AIC/zorinas2017}.
Nevertheless, compared to CT and MRI, 3D Echo is widely available in hospitals at a much lower cost and provides real-time cardiac anatomical and function information for the fusion.
With the recent availability of commercially available fusion packages on commercial X-ray systems, a wider range of clinical applications has been developed based on image fusion related to X-rays.
\Leireffig{fig:method:mm-fusion application} presents a clinical application example of X-ray and Echo image fusion.

\subsubsection{Multi-sequence MRI fusion} \label{method:fusion:MS-CMR}
As MRI has different imaging sequences, one could fuse multi-sequence MRI to combine complementary information from different sequences.
For example, \citet{conf/ISBI/cordero2012} fused cine and LGE MRIs for precise representation of both the Myo boundaries and scars.
\citet{journal/Cirir/zhang2021} fused cine and T1 mapping MRIs to generate gadolinium-free LGE-like MRI via a DL-based generator.
Reducing the need for gadolinium can dramatically shorten scan times, lower the cost of associated consumables, minimize patient preparation time, and eliminate the need for physician presence.
T1 mapping appears the most promising gadolinium-free technique but its clinical utility is still limited by the lack of standardized interpretation and post-processing.
Therefore, one could combine T1 mapping and cine MRIs to exploit and enhance existing contrast and signals within them and display them in a standardized presentation. 

\subsubsection{Discussion}
Image fusion typically includes two stages: (a) image registration (please refer to Section \ref{tb:method:registration}); (b) fusion of relevant features, which could be gray-level based or component-based analysis, from the aligned images.
To avoid the registration step, several hybrid scanners have been developed to integrate multiple modalities \citep{journal/IJR/sazonova2017,journal/CGF/lawonn2018}.
With the scanners, there still exist registration errors mainly due to cardiac or respiratory motion for cardiac imaging, so a post-alignment process remains desired \citep{journal/IJC/garcia2009}.
The feature fusion involves the identification and selection of the features for specific clinical assessment purposes.
In this study, we have found that the most common hybrid version is combining low-resolution molecular images (PET/ SPECT) with anatomical context from high-resolution MR/ CT; 
The most common target regions of fusion are LV Myo and LV, which are mainly involved with the management of CAD and the lead implantation for CRT, respectively.

\begin{figure*}[t]\center
 \includegraphics[width=0.9\textwidth]{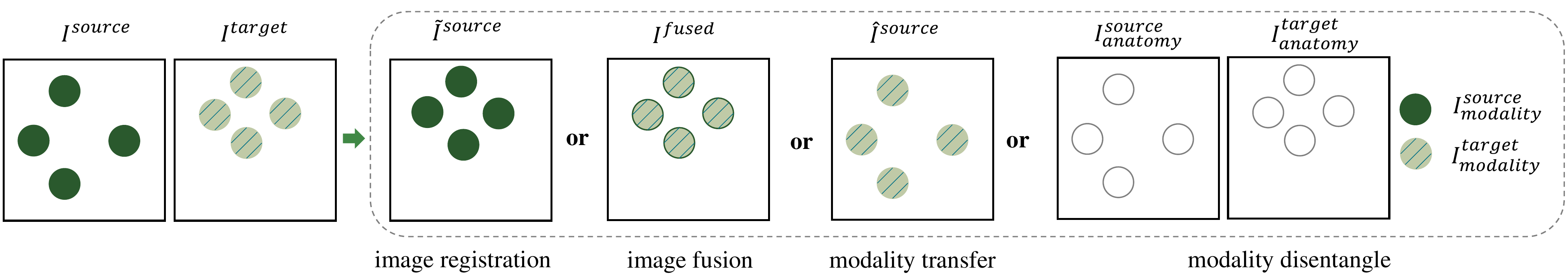}\\[-2ex]
   \caption{Four typical multi-modality image processing schemes. Here, $I^{target}$ denotes target image, while $I^{source}$ refers to source image. $\widetilde{I}^{source}$, $I^{fused}$, $\hat{I}^{source}$ are deformed source image, fused image, and target-style source image, respectively. 
   $I_{anatomy}$ and $I_{modality}$ are the disentangled anatomy image and modality information, respectively.
   Note that in this illustration the modality fusion is performed on the pre-aligned images, i.e., $\widetilde{I}^{source}$ and $I^{target}$.
   }
\label{fig:method:mm-seg}
\end{figure*}

Compared to other medical image fusion (brain, breast, liver, etc.) where many effective fusion schemes have been developed, cardiac image fusion is still at an early stage \citep{journal/IF/james2014}.
The works we surveyed did not provide many details on the fusion schemes, and most of them achieved static fusion.
There were only a few software products commercially available for real-time fusion between dynamic images.
For example, EchoNavigator is a software tool that enables real-time image synchronization and fusion of 2D and 3D TEE with fluoroscopy images.
The coordinate system of the two imaging modalities can be aligned based on the localization and tracking of the TEE probe.
This software might facilitate complex percutaneous procedures and decrease procedure length as well as radiation dose.
Note that the software was installed only in a few hospitals at the time of publication \citep{journal/CCIR/biaggi2015}, which limits the clinical evaluation of fusion images.
Recently, many DL-based image fusion methods have been developed for brain image fusion \citep{journal/IVC/zhang2021}, but they have not been applied to cardiac image fusion yet.
Nevertheless, there are several DL-based works that achieved implicit cardiac image fusion in the latent space for multi-modality image segmentation (please c.f. Section \ref{method:segmentation:fusion} for details).
Given a large amount of data, deep neural networks are powerful for learning hierarchical feature representations.
Note that there are no general criteria for network architecture design, and the questions of what, when and how, applied to cardiac image fusion,  remain unresolved.
Moreover, current research mainly focuses on the fusion of two modalities, and the fusion of three or more modalities is rarely studied in literature \citep{journal/CSTT/tirupal2021}.
Despite extensive research and test applications, multi-modality image fusion techniques have not yet been extensively translated into clinical routine.
In conclusion, fused imaging further increases the variety of cardiac imaging modalities (modality augmentation), but effective techniques are still under development.



\subsection{Segmentation} \label{method:segmentation}
Multi-modality cardiac image segmentation has received a substantial research attention, including both anatomical segmentation \citep{journal/MedIA/zhuang2020} and pathology segmentation \citep{book/MyoPS/zhuang2020}.
It needs to solve the difference in data distribution among different modalities normally via registration, fusion, or domain adaptation (modality transfer or disentanglement).
\Leireffig{fig:method:mm-seg} provides the sketch maps of these modality processing schemes.
The image registration employs spatial transformation parameters to align images of different modalities and then performs segmentation on the anatomy-aligned images \citep{conf/MICCAI/luo2020,journal/TPAMI/Zhuang2019}.
Based on the registration results, one could further fuse several modalities into a single one, and then perform segmentation on the integrated images/ features \citep{journal/MedIA/zhuang2016}.
Moreover, one could utilize transfer learning to transfer different modalities to the target domain and then perform segmentation on the modality-aligned images/ features \citep{conf/STACOM/chen2019}.
Another similar idea is to disentangle the anatomy and modality information from cardiac images, and the target region can be directly segmented on the disentangled anatomy images/ features \citep{journal/MedIA/chartsias2019}.
\Leireftb{tb:method:segmentation} summarizes the representative multi-modality cardiac image segmentation methods, the applied modalities and the target substructures of the heart.

\begin{table*} [htbp] \center
    \caption{Summary of previously published representative works on \textit{multi-modality cardiac segmentation}.
    MM-WHS: unpaired cardiac CT and MRI dataset from Multi-Modality Whole Heart Segmentation Challenge \citep{journal/MedIA/zhuang2020};
    MyoPS: paired cardiac bSSFP, T2-weighed, and LGE MRIs from Myocardial Pathology Segmentation Combining Multi-sequence CMR Challenge \citep{journal/MedIA/li2022-myops};
    X-rayAI: X-ray angiography image; T1-CE: contrast-enhanced T1 imaging; 
    MAS: multi-atlas segmentation; CNN: Convolutional neural network; MTL: multi-task learning; HM: histogram matching; UDA: unsupervised domain adaptation; FCNN: full convolutional neural network; MMD: maximum mean discrepancy;
    Reg-based seg: registration based segmentation;
    $^\dagger$ denotes that the evaluation dataset also includes non-cardiac or single-modality dataset which is out of the scope of this study. 
     }
\label{tb:method:segmentation}
{\small
\scalebox{0.88}
{
\begin{tabular}{p{0.1cm}p{3.3cm}|p{3.6cm}p{3cm}p{5.8cm}p{2.3cm}}
\hline
& Study & Modality & Target & Method & Type\\
\hline
\multirow{7}{*}{\rotatebox{90}{Reg-based seg}} & \citet{journal/MedIA/peters2010}    & CT, MRI, X-rayAI  & WH; LA               & simulated search                         & supervised \\
& \citet{journal/MedIA/zhuang2016}    & CT, MRI           & WH                   & multi-scale and multi-modality MAS       & supervised \\
& \citet{conf/STACOM/zheng2019}       & MS-CMRSeg         & LV, RV, Myo          & DL-based registration and segmentation   & supervised \\
& \citet{conf/MICCAI/ding2020}        & MM-WHS            & Myo                  & DL-based cross-modality MAS              & supervised \\
& \citet{journal/TPAMI/Zhuang2019}    & MS-CMRSeg         & Myo                  & multivariate mixture model               & supervised \\
& \citet{conf/MICCAI/luo2020}         & MS-CMRSeg         & LV, RV, Myo          & DL-based multivariate mixture model      & supervised \\
& \citet{journal/CMPB/paknezhad2020}  & tagged, cine MRI  & Myo                  & deformation-based segmentation           & supervised \\
\hline\hline
\multirow{9}{*}{\rotatebox{90}{Fusion based segmentation}} & \citep{conf/STACOM/mortazi2017}     & MM-WHS            & WH                   & multi-planar network with an adaptive fusion strategy  & supervised \\
& \citet{conf/STACOM/tong2017}        & MM-WHS            & WH                   & modality normalization and fusion        & supervised \\
& \citet{conf/MyoPS/zhao2020}         & MyoPS             & LV scar and edema    & stacked and parallel U-Nets for fusion   & supervised \\ 
& \citet{conf/MyoPS/jiang2020}        & MyoPS             & LV scar and edema    & max-fusion U-net                         & supervised \\ 
& \citet{journal/BSPC/li2022}         & MS-CMRSeg, MyoPS  & LV, RV, Myo; LV scar and edema    & deep co-training            & supervised \\ 
& \citet{conf/MyoPS/ankenbrand2020}   & MyoPS             & LV scar and edema    & ensemble U-net                           & supervised \\ 
& \citet{conf/MyoPS/zhai2020}         & MyoPS             & LV scar and edema    & coarse-to-fine weighted ensemble model   & supervised \\ 
& \citet{conf/MyoPS/martin2020}       & MyoPS             & LV scar and edema    & stacked BCDU-net with MRI synthesis      & supervised \\ 
& \citet{journal/MedIA/wang2022}      & MyoPS             & LV scar and edema    & auto-weighted attention ensemble model   & supervised \\ 
\hline \hline
\multirow{35}{*}{\rotatebox{90}{Domain adaptation based segmentation}} & \citet{conf/CVPR/zhang2018}         & CT, MRI           & WH                   & cycle- and shape-consistency GAN         & supervised \\
& \citet{conf/AAAI/chen2019}          & MM-WHS            & LA, LV, Myo, AO      & synergistic image and feature adaptation & unsupervised \\ 
& \citet{conf/STACOM/tao2019}         & MS-CMRSeg         & LV, RV, Myo          & shape-transfer GAN                       & unsupervised \\
& \citet{conf/STACOM/chen2019}        & MS-CMRSeg         & LV, RV, Myo          & image-to-image translation               & unsupervised \\
& \citet{conf/STACOM/ly2019}          & MS-CMRSeg         & LV, RV, Myo          & style data augmentation                  & unsupervised \\
& \citet{conf/STACOM/wang2019}        & MS-CMRSeg         & LV, RV, Myo          & HM and domain adversarial learning       & unsupervised \\
& \citet{journal/IAccess/dou2019}     & MM-WHS            & LA, LV, Myo, AO      & plug-and-play adversarial UDA            & unsupervised \\ 
& \citet{conf/MICCAI/ouyang2019}      & MM-WHS            & LA, LV, Myo, AO      & VAE-based feature prior matching         & unsupervised \\ 
& \citet{conf/MICCAI/chen2020}        & MM-WHS, MS-CMRSeg & LA, LV, RV, Myo, AO  & affinity-guide CNN                        & supervised \\
& \citet{journal/CMIG/liao2020}       & MM-WHS            & WH                   & multi-modality transfer learning         & semi-supervised \\
& \citet{journal/TNNLS/yu2020}        & CT, MRI           & Myo                  & MTL + adversarial reverse mapping        & supervised \\
& \citet{conf/PRCV/liu2020}           & MS-CMRSeg         & LV, RV, Myo          & cycleGAN with MMD constraints            & unsupervised \\ 
& \citet{journal/TMI/wu2020}          & MM-WHS, MS-CMRSeg & LV, RV, Myo          & explicit domain discrepancy              & unsupervised \\
& \citet{journal/TMI/li2020}          & MM-WHS            & WH                   & dual-teacher++                           & semi-supervised \\
& \citet{conf/MICCAI/xue2020}         & MM-WHS            & WH                   & dual-task and hierarchical learning      & unsupervised \\
& \citet{journal/MedIA/bian2020}      & MM-WHS$^\dagger$  & LA, LV, Myo, AO      & uncertainty-aware domain alignment       & unsupervised \\ 
& \citet{journal/TMI/wu2021}          & MM-WHS, MS-CMRSeg & WH; LV, RV, Myo      & variational approximation                & unsupervised \\ 
& \citet{journal/APM/wang2021}        & MM-WHS            & LA, LV, Myo, AO      & global and category-wise alignment       & unsupervised \\ 
& \citet{journal/TMI/bian2021}        & MM-WHS$^\dagger$  & LA, LV, Myo, AO      & zero-shot learning                       & unsupervised \\ 
& \citet{journal/TMI/chen2021}        & MM-WHS$^\dagger$  & LA, LV, Myo, AO      & information bottleneck GAN               & unsupervised \\ 
& \citet{journal/MedIA/chen2021}      & MM-WHS$^\dagger$  & LA, LV, Myo, AO      & diverse data augmentation GAN            & unsupervised \\ 
& \citet{journal/CBM/cui2021}         & MM-WHS            & LA, LV, Myo, AO      & bidirectional UDA                        & unsupervised \\ 
& \citet{journal/TMI/cui2021}         & MM-WHS            & LA, LV, Myo, AO      & hybrid domain-invariant information      & unsupervised \\ 
& \citet{journal/TMI/tomar2021}       & MM-WHS$^\dagger$  & LA, LV, Myo, AO      & spatial adaptive normalization           & unsupervised \\
& \citet{journal/TMI/vesal2021}       & MM-WHS, MS-CMRSeg & LV, RV, Myo          & point-cloud shape adaptation             & unsupervised \\ 
& \citet{conf/MICCAI/zeng2021}        & MM-WHS$^\dagger$  & LA, LV, Myo, AO      & semantic consistent UDA                  & unsupervised \\ 
& \citet{journal/MedIA/guo2021}       & MRI, CT, Echo     & LV, Myo              & few-shot multi-level semantic adaptation & semi-supervised \\
& \citet{journal/MedIA/liu2021}       & MM-WHS            & LA, LV, Myo, AO      & symmetric FCNN with attention            & unsupervised \\
& \citet{conf/ISCTCC/kots2021}        & MM-WHS$^\dagger$  & LA, LV, Myo, AO      & deep co-training                         & semi-supervised \\
& \citet{journal/TMI/cai2016}         & CT, MRI           & LA, LV, Myo; WH      & groupwise; spectral decomposition        & unsupervised \\
& \citet{journal/TMI/dou2020}         & MM-WHS$^\dagger$  & LA, LV, Myo, AO      & knowledge distillation                   & supervised \\
& \citet{conf/STACOM/chartsias2019}   & bSSFP + LGE MRI   & LV, Myo              & disentangled representation learning     & semi-supervised \\
& \citet{journal/MedIA/chartsias2019} & MM-WHS$^\dagger$  & WH                   & disentangled representation learning      & semi-supervised \\
& \citet{journal/TMI/chartsias2020}   & bSSFP, LGE MRI$^\dagger$ & LV, Myo       & disentangle align and fuse network       & semi-supervised \\ 
& \citet{journal/MedIA/pei2021}       & MM-WHS, MS-CMRSeg & LV, RV, Myo          & disentangle domain features              & unsupervised \\ 
& \citet{journal/MedIA/wang2021}      & MM-WHS, MS-CMRSeg & LV, RV, Myo          & cycle-consistent model                   & unsupervised \\ 
\hline
\end{tabular}}
}
\end{table*}

\subsubsection{Registration based segmentation}  \label{method:segmentation:registration}
A direct way to propagate anatomy knowledge between modalities is multi-modality registration, and these methods can be categorized as registration based segmentation.
Among registration based segmentation methods, atlas registration based approaches are commonly used in multi-modality scenarios.
They register one or multiple atlases to the target image, followed by propagation of (usually manual) labels.  
We refer the reader to Sec. \ref{tb:method:registration} for multi-modality registration.
If several atlases are available, labels from individual atlases can be combined into a final segmentation via a label fusion strategy \citep{journal/MedIA/zhuang2016,conf/MICCAI/ding2020}.
The number of atlases determines the potential optimal performance of multi-atlas segmentation (MAS), and thus considering multi-modality atlases is beneficial when they are available.
Conventional multi-modality MAS based methods are generally computationally expensive. 
This is because these methods perform the registration step in an iterative fashion and typically employ patch-based label fusion \citep{journal/TMI/bai2013,journal/MedIA/zhuang2016,journal/MedIA/sanroma2018}. 
To solve this, one could achieve both image registration and label fusion by deep neural networks for a computationally efficient MAS framework \citep{conf/MICCAI/ding2020}.	

However, these DL-based MAS frameworks can not be optimized in an end-to-end fashion, as the image registration and label fusion are separated into two tasks.
Recently, \citet{conf/MICCAI/luo2020} proposed a probabilistic image registration framework based on a multivariate mixture model and neural network estimation.
In their framework, MAS was unified by groupwise registration, and registration and segmentation of multi-modality cardiac images were achieved simultaneously.
The joint distribution of multi-modality images was modeled as multivariant mixtures, and the model was formulated with transformation.
Segmentation can be performed on a virtual common space, where all the images were simultaneously registered.
Instead of the commonly used Expectation Maximization (EM) algorithm  \citep{journal/TPAMI/Zhuang2019}, they utilized neural networks to efficiently estimate the parameters of multivariate mixture model \citep{conf/MICCAI/luo2020}.

\begin{figure*}[t]\center
 \includegraphics[width=1\textwidth]{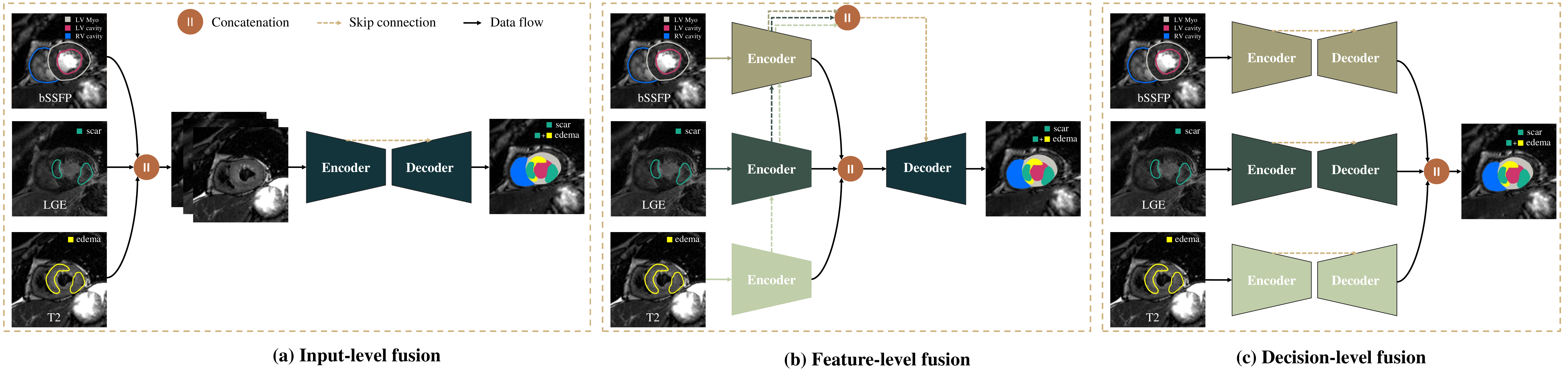}\\[-2ex]
   \caption{An example of multi-modality cardiac image segmentation. Here, we present three different deep learning based fusion strategies for myocardial pathology segmentation combining multi-sequence MRIs. Note that here bSSFP, T2-weighed, and LGE MRIs have been aligned before fusion.}
\label{fig:method:mm-seg:fusion stage}
\end{figure*}

\subsubsection{Fusion based segmentation}  \label{method:segmentation:fusion} 
To the best of our knowledge, there are only a few works that employ image fusion for multi-modality cardiac segmentation.
Most fusion based segmentation methods do not explicitly generate a new integrated image by image fusion (also namely \textit{explicit fusion}), but only perform an \textit{implicit fusion} at different levels.
The only available explicit fusion based segmentation, up to our knowledge, is from \citet{conf/STACOM/tong2017}, where cardiac MRI and CT were fused based on their normalized mean intensity values.
With the fused images, the size of the training set was increased, and therefore the segmentation model yielded better results.
With the popularization of DL methods, DL-based image fusion has emerged, as it can learn high-level features from data in an incremental manner compared with conventional methods \citep{journal/Array/zhou2019}.
It mitigates the need for domain expertise, simplifies the feature extraction steps, and can be optimized in an end-to-end style.

\Leireffig{fig:method:mm-seg:fusion stage} presents the three typical types of DL-based modality fusion for multi-modality cardiac image segmentation.
Many works simply concatenate different modalities into different channels and focus on the subsequent segmentation network architecture, namely input-level fusion \citep{conf/STACOM/wangXY2019,conf/MyoPS/yu2020,conf/MyoPS/zhangXR2020,conf/MyoPS/elif2020}.
Similarly, one could perform fusion by fusing the model outputs from different modality inputs, namely decision-level fusion.
For the decision-level fusion, many fusion strategies can be used, such as averaging and majority voting \citep{journal/AIR/rokach2010}.
Both input- and decision-level fusion are simple, but ignore the relationship among different modalities.
A potential solution is to utilize two-stage networks which consider the prior spatial relationship between different modalities.
One typical application is  MyoPS, where anatomical information can be extracted from one modality (bSSFP MRI) at the first stage;
with obtained segmentation masks from the first stage, the functional information can be learned by combining other modalities (LGE and T2 MRIs) at the second stage \citep{book/MyoPS/zhuang2020}.
Moreover, fusion can be performed on latent feature space \citep{conf/STACOM/mortazi2017,conf/MyoPS/jiang2020,journal/BSPC/li2022,conf/MyoPS/zhao2020}, namely feature-level fusion, and multi-scale information fusion can be achieved via densely connected layers \citep{conf/MyoPS/zhao2020}.
For example, \citet{conf/MyoPS/jiang2020} proposed a max-fusion U-Net for multi-sequence cardiac pathology segmentation, where the features from different modalities were fused via a pixel-wise maximum operator.
The operator was expected to guide the network to keep informative features extracted by each modality.
Furthermore, the contribution of each modality can be weighted via cross-modal convolutions \citep{conf/CVPR/tseng2017}.

Compared with input/ decision-level fusion, feature-level fusion can pay more attention to learning the complex relationship between different modalities \citep{journal/Array/zhou2019}.
However, the features to be merged are not modality-invariant, and therefore may affect the multi-modality segmentation.
To solve this, one could combine disentangled representation learning with fusion schemes for the segmentation.
For example, \citet{journal/TMI/chartsias2020} disentangled modality and anatomy features from multi-modality images, and then fused the disentangled anatomy features for cardiac segmentation of bSSFP and LGE MRIs.
Besides modality fusion, model ensemble scheme is also widely employed for multi-modality cardiac segmentation by integrating several predictions from different models \citep{conf/MyoPS/ankenbrand2020,conf/MyoPS/zhai2020,conf/MyoPS/martin2020}.

\subsubsection{Domain adaptation based segmentation}  \label{method:segmentation:DA}
\begin{figure}[t]\center
 \includegraphics[width=0.49\textwidth]{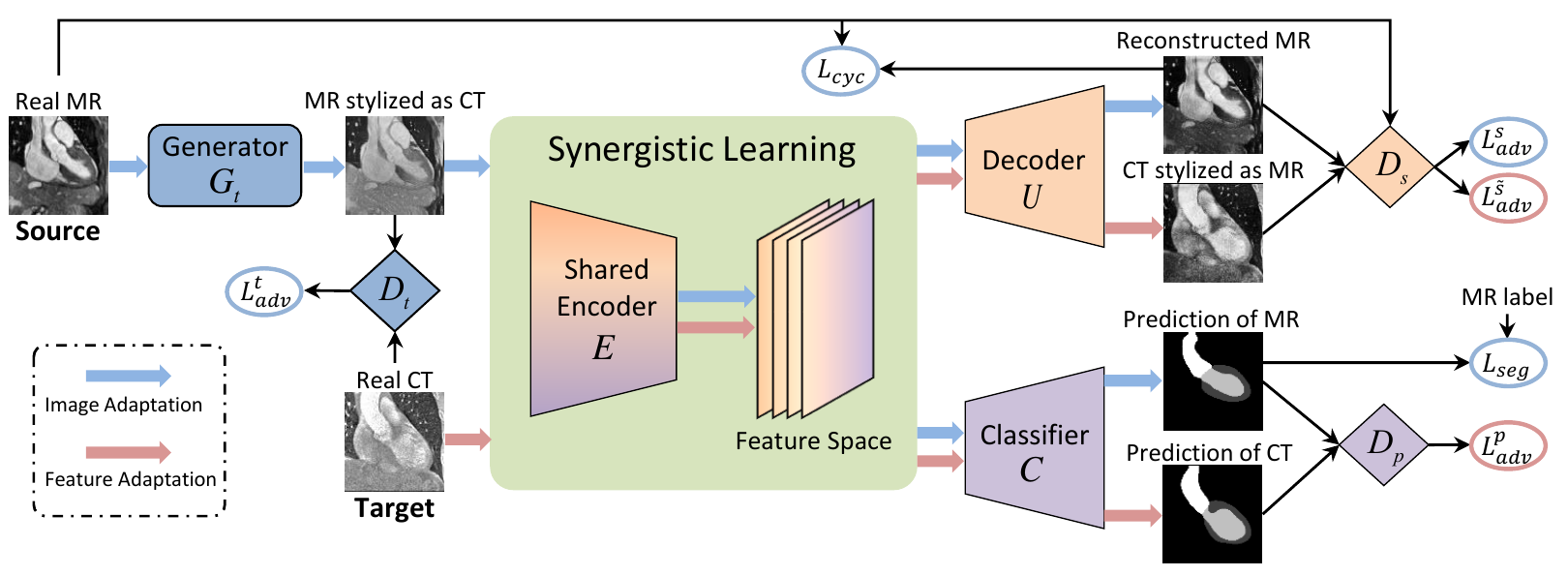}\\[-2ex]
   \caption{An example of unsupervised domain adaptation based cross-modality cardiac image segmentation. Image adapted from \citet{conf/AAAI/chen2019} with permission.}
\label{fig:method:mm-seg:domain adaptation}
\end{figure}

For multi-modality cardiac analysis, the annotation of all modalities via supervised learning can be tedious, time-consuming, and significantly variable across imaging modalities.
An intuitive solution is to transfer the knowledge from the annotation-rich modalities (denoted as source domain) to another annotation-poor modality (referred to as target domain).
Note that here the source and target domains could be collected from the same or different subjects.
Due to the existence of domain shift, the model trained on the source domain usually fails on the target domain \citep{journal/TMI/wu2020,journal/MedIA/li2022}.
Domain adaptation is therefore proposed to generalize models from the source domain to the target domain without much performance degradation.
This is normally achieved by aligning two domains into a common space, where their domain discrepancy can be minimized.
Regarding the manner of domain alignments, current domain adaptation can be categorized into three kinds, i.e., discrepancy minimization-based strategies, adversarial learning-based algorithms and reconstruction-based methods \citep{journal/MedIA/liu2021}.
There are many applications of domain adaptation, such as semantic segmentation, image classification, and object detection.
Specific to the cardiology field, domain adaptation has been employed for cross-modality image segmentation \citep{journal/MedIA/pei2021}, cross-axis MRI segmentation \citep{journal/TMI/koehler2021}, cardiac strain analysis of Echo \citep{journal/TMI/lu2021}, cross-individual ECG arrhythmia classification \citep{conf/EMBC/chen2020}, and comparison of cardiac simulation models \citep{conf/FIMH/duchateau2019}.
We mainly summarize the work of cross-modality cardiac image segmentation via domain adaptation, either by modality transfer or modality disentangle, as presented in the bottom part of \Leireftb{tb:method:segmentation}. 
One can see cardiac cross-modality domain adaptation works emerge from 2019, with the release of two public cardiac multi-modality/ sequence datasets.
All these works are DL-based and are generally unsupervised, i.e., without using the target domain label.
\Leireffig{fig:method:mm-seg:domain adaptation} presents an example where unsupervised domain adaptation is applied to segment cardiac CT with the assistance of MRI.
Two unique semi-supervised works are based on dual-teacher++ \citep{journal/TMI/li2020} and few shot \citep{journal/MedIA/guo2021}, which both employed only limited target domain label.

\subsubsection{Discussion}  \label{method:segmentation:summary}
Multi-modality image segmentation benefits from both aforementioned image registration and fusion especially when involving a combination of multiple modalities.
However, currently image registration or explicit image fusion is mostly used as a pre-processing step for multi-modality segmentation \citep{journal/MedIA/li2022-myops,conf/STACOM/tong2017}. 
Also, no work is able to simultaneously achieve multi-modality fusion from unregistered data for cardiac image segmentation, as far as we know.
Only a few studies achieved simultaneous registration and segmentation for multi-modality cardiac images via multivariate mixture models \citep{journal/TPAMI/Zhuang2019,conf/MICCAI/luo2020}.
Unified registration and segmentation frameworks eliminate the need for a separate registration model, and segmentation can facilitate registration and vice versa.
In the future, it would be desirable to achieve simultaneous registration and fusion of multi-modality data for cardiac segmentation.

Domain adaptation can be regarded as a new application of multi-modality imaging.
It does not combine information from different modalities but transfers information across modalities, to assist the image segmentation of the target domain which usually has limited available annotated data.
Modality information transfer can also be employed in cardiac image reconstruction, where cardiac motion priors can be obtained from other modalities for motion correction \citep{journal/TBME/sang2021}. 
An alternative idea to domain adaptation that appears in multi-modality image registration is the disentangled representation.
It is reasonable as both multi-modality image registration and segmentation care about the structural information instead of the modality information.
Moreover, compared to cardiac image registration and fusion, DL-based models have been widely employed for multi-modality cardiac image segmentation, thanks to the release of public datasets. 


\begin{table*} [htbp] \center
    \caption{Public multi-modality cardiac image datasets. 
     }
    \label{tb:evaluation:data}
    {\small
    \begin{tabular}{l| llll}
    \hline
    Reference    & Dataset	& Target  & Disease  &  Type \\
    \hline
    \citet{link/CARMA2012}            & 155 bSSFP, LGE MRI	   & LA cavity and scar  & AF              & paired \\
    \citet{journal/MedIA/tobon2013}   & 15 MRI, 3D Echo        & Myo                 & healthy         & paired \\
    \citet{journal/MedIA/tobon2015}   & 30 MRI, 30 CT          & LA cavity           & AF              & unpaired \\
    \citet{journal/MedIA/karim2018}   & 10 MRI, 10 CT          & LA wall             & AF              & unpaired \\
    \citet{journal/MedIA/zhuang2019}  & 60 MRI, 60 CT          & WH                  & AF, CHD, others & unpaired \\
    \citet{journal/MedIA/zhuang2020}  & 45 bSSFP, T2 weighted, LGE MRI  & LV, RV, Myo         & MyoI   & paired \\
    \citet{journal/MedIA/li2022-myops}& 45 bSSFP, T2 weighted, LGE MRI  & LV scar and edema   & MyoI   & paired \\
    \hline
\end{tabular}}\\
\end{table*}

\section{Data and evaluation measures} \label{cardiac data and evaluation}

\subsection{Public multi-modality cardiac datasets} \label{public dataset}
\href{https://www.cardiacatlas.org/}{ The Cardiac Atlas Project} provides a public imaging database for computational modeling and statistical atlases of the heart \citep{journal/BI/fonseca2011}.
It includes  images targeted to different cardiac structures as well as diseases and acquired with different modalities from different centers.
Other challenge events have released public multi-modality cardiac datasets, as summarized in \Leireftb{tb:evaluation:data}.
One can see that different from the computer vision datasets, there are a limited number of public multi-modality cardiac image datasets.
Also, only some specific modalities are covered, including MRI, CT and Echo.
Nevertheless, these datasets have already substantially facilitated the development of multi-modality cardiac imaging analysis, as we summarized in Sec \ref{method:segmentation}.
In the future, more multi-modality cardiac images, either paired or unpaired, are expected to be released to boost the development of multi-modality cardiac image analysis.   

\subsection{Evaluation measures} \label{evaluation measures}
Thorough validation is necessary for clinical acceptance.
Verifying the accuracy of multi-modality image registration or fusion is a difficult task as a good quality gold standard is normally unavailable.
Visual assessment, sometimes used to evaluate the accuracy, is subjective and thus not reproducible. 
For an objective evaluation, external labels, anatomical landmarks and/or external fiducial frames are required  \citep{journal/TMI/makela2002}.
In contrast, the evaluation of multi-modality image segmentation is relatively simple as manual segmentation can be used as the gold standard.
Nevertheless, the comparison of measures from literature is difficult as different metrics can be applied to evaluate one method.
Note that sometimes different evaluation metrics could lead to different conclusions regarding the performance of an algorithm, indicating the potential limitation of current metrics.
In this section, we summarize several common measures employed in each multi-modality image computing task.

\subsubsection{Image registration measures} \label{evaluation measures:registration}
\begin{figure}[t]\center
 \includegraphics[width=0.49\textwidth]{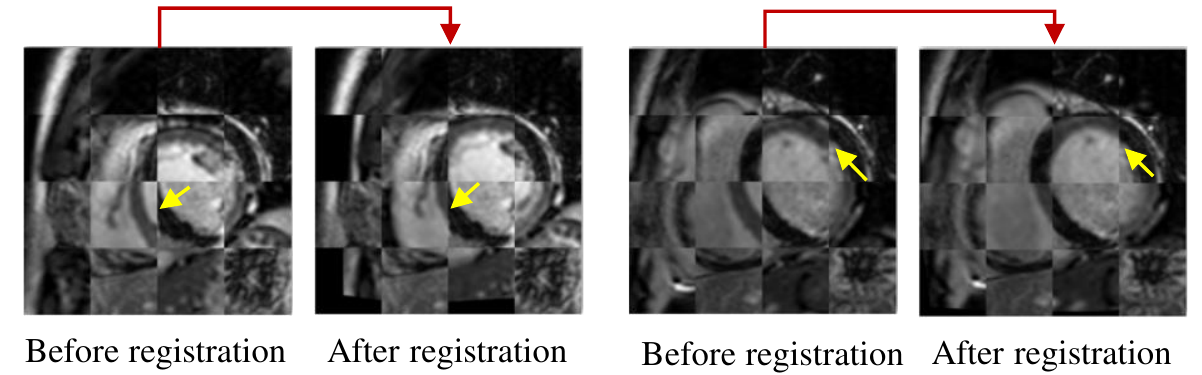}\\[-2ex]
   \caption{Two examples of multi-modality cardiac image registration (bSSFP and LGE MRIs) visual evaluation via overlaid blocks. The representative visible fixed misalignment areas are highlighted by the yellow arrows. }
\label{fig:evaluation:reg:overlaid block}
\end{figure}

For assessing the performance of cardiac multi-modality image registration, a range of different measures have been explored \citep{journal/TMI/van2005}.
The registration may involve different combinations of dimensions, with the most common combinations being 3D-to-3D and 2D-to-3D.
For 3D-to-3D registration, one could employ mean landmark distance (mLD) \citep{journal/AJR/sinha1995}, mean target registration error (mTRE), and surface distance \citep{journal/JCI/sturm2003,conf/MICCAI/makela2001} for evaluation.
Here, mLD can be defined as,
\begin{equation}
    \operatorname{mLD}\left(\hat{T}\right)=\frac{1}{N_l} \sum_{i,j=1}^{N_\mathbf{l}}\left\|\hat{T}(\mathbf{l}_{i})- \mathbf{l}_{j}\right\|,
\end{equation}
where $\hat{T}$ is the predicted transformation, $N_l$ is the number of landmarks, and $\mathbf{l}_{i}$ and $\mathbf{l}_{j}$ are paired landmarks from source and target images, respectively.
mTRE is computed as the mean 3D distance between the point $\mathbf{p}_{j}$ from the target image space and the corresponding point $\mathbf{p}_{i}$ from the source image space transformed with the predicted transformation $\hat{T}$,
\begin{equation}
    \operatorname{mTRE}\left(\hat{T}\right)=\frac{1}{N_p} \sum_{i,j=1}^{N_p}\left\|\hat{T}(\mathbf{p}_{i})-\mathbf{p}_{j}\right\|,
\end{equation}
where $N_p$ is the number of points.
mTRE is a good measure for patient-to-image registration for navigation, as it calculates the misalignment on specific points of interest within the target volume where the operation will be performed.
Surface distance can be defined as the mean 3D Euclidian distance, standard deviation, and root mean square distance between the transformed source image and target image surfaces \citep{journal/JCI/sturm2003}.
Note that here the surface is sampled after registration in the format of paired corresponding points.
Therefore, one could employ the percent number of overlapped point pairs as a registration measure.
Checkerboard visualization for visual assessment is also commonly used \citep{journal/MedIA/makela2003,conf/FIMH/camara2009,conf/ISBI/courtial2019}, as presented in \Leireffig{fig:evaluation:reg:overlaid block}.
Due to the absence of the gold standard, one could instead evaluate the anatomical labels of transformed source images in terms of Dice and Hausdorff distance (HD) \citep{journal/TNS/turco2016,journal/JCARS/atehortua2020}, defined as,
\begin{equation}
  \mathrm{Dice}(V_{\mathrm{1}}, V_{\mathrm{2}}) = \frac{2\left|V_{\mathrm{1}} \cap V_{\mathrm{2}}\right|}{\left|V_{\mathrm{1}}\right|+\left|V_{\mathrm{2}}\right|},
\end{equation}
and
\begin{equation}
  \mathrm{HD}(X, Y)=\max \Big[\sup _{x \in X} \inf _{y \in Y} d(x, y), \sup _{y \in Y} \inf _{x \in X} d(x, y)\Big],
\end{equation}
where $V_{\mathrm{1}}$ and $V_{\mathrm{2}}$ denote the set of pixels in the target and transformed source labels, respectively;
$|\cdot|$ refers to the number of pixels in set $V$;
$X$ and $Y$ represent two sets of contour points;
and $d(x, y)$ indicates the Euclidean distance between the two points $x$ and $y$;
Besides, one could randomly transform the source image with known transformations for the known gold standard to quantify registration accuracy \citep{conf/FIMH/pauna2003}.

For 2D-to-3D registration, visual inspection \citep{journal/MP/aksoy2013}, mTRE$_{\text{proj}}$ \citep{journal/CMBBE/gouveia2017}, and centreline 2D distance measure \citep{journal/MedIA/baka2013,journal/TMI/van2005} have been utilized for evaluation.
Here, mTRE$_{\text{proj}}$ refers to the mTRE in the projection direction and is defined as,
\begin{equation}
    \operatorname{mTRE_{\text{proj}}}\left(\hat{T}\right)=\frac{1}{N_p} \sum_{i,j=1}^{N_p}\left\|(\hat{T}(\mathbf{p}_{i})-\mathbf{p}_{j}) \cdot \hat{\mathbf{n}}\right\|,
\end{equation}
where $\hat{\mathbf{n}}$ is the normal to the projection plane.
However, 2D error measures may be more appropriate than mTRE, as 2D-to-3D registration normally projects a 3D image onto a 2D image plane.
Another solution is employing mean reprojection distance (mRPD), which calculates the minimum distance between the line (eg. centreline of CA) from 2D projected points to the corresponding 3D position of that point.
Here, mRPD can be defined as,
\begin{equation}
    \operatorname{mRPD}\left(\hat{T}\right)=\frac{1}{N_p} \sum_{i,j=1}^{N_p}\text{Dist}({L}_{i}(\hat{T}(source, \mathbf{p}_{i})), \mathbf{p}_{j}),
\end{equation}
where $\text{Dist}({L}_{i}, \mathbf{p}_{j})$ denotes the minimum distance between the 3D point $\mathbf{p}_{j}$ and a line ${L}_{i}$ through $\mathbf{p}_{i}$ and the source image (2D image) space.

Besides, the Jacobian determinant can measure the plausibility of estimated displacement fields for non-rigid registration \citep{journal/NI/ashburner2007}. 
The Jacobian matrix of a displacement field $\mathbf{\Phi}$ at a spatial point $\mathbf{p}$ is defined as,
\begin{equation}
    J_{\mathbf{\Phi}}(\mathbf{p})=\nabla \mathbf{\Phi}(\mathbf{p})
\end{equation} 
where $\mathbf{\Phi}(\mathbf{p})$ denotes the displacement value at $\mathbf{p}$. 
Jacobian determinant $\det(J_{\mathbf{\Phi}}(\mathbf{p})) \le 0$ indicates that $\mathbf{\Phi}(\mathbf{p})$ is not topology-preserving, while the values near 1 represent smooth fields \citep{conf/NIPS/dalca2019}. 

\subsubsection{Image fusion measures} \label{evaluation measures:fusion}

Although there are several multi-modality cardiac image fusion works, most of them are performed in clinical studies generally with visual evaluation \citep{journal/IPADCD/chauhan2021}.
There are many available objective evaluation measures for image fusion, including entropy, MI, standard deviation (SD), peak signal to noise ratio (PSNR), structural similarity index measure, average gradient, Q$_\text{AB/ F}$metric, root mean square error (RMSE), edge intensity, visual information fidelity, spatial frequency, and spectral discrepancy \citep{journal/CSTT/tirupal2021,journal/PCS/bhavana2015}.
This qualitative analysis mainly aims to check the quality of fused images from different perspectives.
However, specific to the cardiology field, the measure of image fusion is mainly performed at the registration stage instead of the fusion stage, as to our knowledge.
This may be due, at least in part, to the limited methodology development of image fusion in cardiology.

\subsubsection{Image segmentation measures} \label{evaluation measures:segmentation}
The most commonly used measure for multi-modality cardiac image segmentation is the Dice score.
Jaccard index, HD, and average surface distance (ASD) have also been used \citep{journal/JBHI/li2021,journal/KBS/li2021,journal/TPAMI/Zhuang2019}, defined as:
\begin{equation}
  \mathrm{Jaccard}(V_{\mathrm{auto}}, V_{\mathrm{manual}}) = \frac{\left|V_{\mathrm{auto}} \cap V_{\mathrm{manual}}\right|}{\left|V_{\mathrm{auto}}  \cup V_{\mathrm{manual}}\right|},
\end{equation}
and
\begin{equation}
    \mathrm{ASD}(X, Y)=\frac{1}{2}\left(\frac{\sum_{x \in X} \min _{y \in Y}d(x, y)}{\sum_{x \in X} 1}+\frac{\sum_{y \in Y} \min _{x \in X}d(x, y)}{\sum_{y \in Y} 1}\right),
\end{equation}
where $V_{\mathrm{auto}}$ and $V_{\mathrm{manual}}$ denote the set of pixels in the automatic and manual segmentation, respectively.
Statistical measurements can also be employed \citep{journal/JBHI/li2021,conf/STACOM/mortazi2017}, i.e., Accuracy (Acc), Specificity (Spe), Sensitivity (Sen), and Precision (Pre). 
Acc refers to the proportion of pixels that have received the correct label among the total number of subjects examined.
Spe and Sen (also known as Recall) are utilized to reflect the success rates of algorithms for the background and the foreground segmentation, respectively.
Precision (also called positive predictive value) indicates the fraction of relevant instances among the retrieved instances.
These measures can appear in various combinations when evaluating multi-modality cardiac image segmentation, and Dice is usually included \citep{journal/TMI/cai2016,conf/MICCAI/ding2020}.


\begin{table*} [!t] 
    \caption{Summary of potential clinical applications of combining multi-modality cardiac imaging for cardiac analysis.
    PAD: peripheral artery disease; ID: infectious endocarditis; DCM: dilated cardiomyopathy; HCM: hypertrophic cardiomyopathy; CS: cardiac sarcoidosis; DMR: degenerative mitral valve regurgitation; AS: aortic stenosis; MVD: mitral valve disease; AVS: aortic valve stenosis;
    TAVR: trans-aortic valve replacement.
     }
\label{tb:discussion:clinical application}
{\small
\begin{tabular}{p{2.5cm}|p{5.8cm}p{2cm}p{6.4cm}} \hline
Modality & Clinical application & Pathology & Representative study \\
\hline 
CT, MRI              & WH segmentation; LA cavity segmentation; LA wall thickness measurement; cardiac index estimation  & AF, CHD, etc  & \citet{journal/MedIA/zhuang2019,journal/MedIA/tobon2015,journal/MedIA/karim2018,journal/TNNLS/yu2020}   \\
CTA, MRI             & CA quantification                   & CAD           & \citet{journal/JCI/sturm2003}   \\
CTA, SPECT           & Myo/ CA quantification              & CAD, ID       & \citet{journal/EJR/koukouraki2013,journal/JBR/piccinelli2013,journal/IJR/sazonova2017}   \\
CTA, X-ray           & Angioplasty; TAVI guidance          & CAD, AVS      & \citet{journal/MedIA/baka2013,journal/CMBBE/gouveia2017}   \\
MRI, X-ray           & invasive cardiovascular procedures (such as CRT); LV contour detection; target endomyocardial injections   & PAD, CHD, MyoI  & \citet{conf/IPMI/oost2003,journal/CCI/gutierrez2007,journal/CiC/ma2010,journal/MP/faranesh2013,journal/CCI/abu2014,journal/Cir/de2006,journal/MedIA/choi2016} \\
CT, Echo             & LA wall thickness measurement; real-time surgical navigation; ablation/ TAVI/ TAVR guidance   & AF, AS, MVD, CHD    & \citet{journal/MedIA/karim2018,journal/TMI/huang2009,conf/CC/sandoval2013,journal/MBEC/khalil2017,journal/JMI/khalil2017,journal/JOC/watanabe2021}   \\
MRI, Echo            & cardiac motion estimation; Echo acquisition guidance; Myo viability assessment & DCM, HCM, MyoI  & \citet{journal/MedIA/puyol2017,journal/JCARS/atehortua2020,conf/IUS/kiss2013,conf/IUS/kiss2011,journal/MedIA/makela2003}   \\
MRI, PET             & practical MRI acquisition; Myo viability, perfusion and metabolism assessment; cardiac motion correction  & MyoI, CAD, CS  & \citet{journal/JNM/kolbitsch2017,journal/AJR/sinha1995,conf/MICCAI/makela2001,journal/JNC/zandieh2018,journal/PTRSA/polycarpou2021}   \\
X-ray, Echo          & LAA closure procedure guidance; ablation and TAVI guidance   & CHD, AF, SHD           & \citet{conf/MICCAI/housden2012,journal/ACD/hadeed2018,journal/CDT/ebelt2020}   \\
\hline \hline
bSSFP, LGE, T2 weighted MRIs  & LV (scar and edema), RV, Myo segmentation     & MyoI  & \citet{journal/MedIA/zhuang2020,journal/MedIA/li2022-myops,journal/MedIA/wang2022} \\
bSSFP, T2 mapping MRIs        & generate gadolinium-free LGE-like MRI & MyoI  & \citet{journal/Cirir/zhang2021} \\
bSSFP, LGE MRIs               & LA fibrosis and scar segmentation     & AF  & \citet{journal/MedIA/li2020,conf/MICCAI/wu2018} \\
tagged, cine MRIs             & Myo strain analysis; motion estimation & DMR & \citet{journal/CMPB/paknezhad2020,journal/TMI/shi2012} \\
\hline
\end{tabular}}\\
\end{table*}

\section{Discussion and future perspectives} \label{discussion}
The last decade has witnessed an enormous amount of efforts in the specific field of multi-modality image computing for cardiac analysis.
Cardiac hybrid imaging can provide valuable diagnostic and prognostic information for patients with CVDs, compared with side-by-side evaluation from a single imaging modality \citep{journal/CTI/gimelli2013}.
\Leireftb{tb:discussion:clinical application} summarizes the potential clinical applications of the developed computing algorithms.
With the development of hybrid imaging devices, there has been a reduction in complex image processing through software.
This is because the superposition of images using a hybrid device, although it may not be obtained simultaneously, is obtained by successive measurements.
Nevertheless, many techniques have not yet been translated into clinical routine.
This is, at least in part,  attributed to the fact that the pre-processing step, i.e., registration, usually requires intensive manual interaction.
In this discussion, we aim to identify the most important challenges that may drive future research in multi-modality cardiac image analysis.

\subsection{Multi-modality cardiac computing with missing modality} \label{missing modality}
In multi-modality cardiac studies, one or more sub-modalities may be missing due to poor image quality (e.g. imaging artifacts), failed acquisitions or counterindications.
However, current multi-modality learning algorithms generally assume that all modalities are available \citep{journal/arxiv/azad2022}.
Nevertheless, there exist a number of methods to deal with missing modalities, which can be roughly categorized into three types.
Firstly, different models can be designed for every potential missing modality combination, which is complicated and time-consuming.
Secondly, many works attempt to impute or synthesize missing modalities to obtain the complete imaging modalities \citep{conf/MICCAI/havaei2016}.
However, an additional network is normally required for synthesis, and the model might be sensitive to the synthesized image quality.
Thirdly, one could learn a shared feature representation through all sub-modalities and then project it to a single model \citep{journal/TMI/zhou2019}.
The third solution is more efficient than the previous two approaches though it is difficult to learn a shared latent representation.
The most straightforward way is employing a mean function to map all available sub-modalities into a unified representation, though this may lose important information \citep{conf/arXiv/lau2019}.
Feature disentanglement and domain adaptation can also be utilized to generate a shared feature representation for missing modalities \citep{conf/MICCAI/chen2019,conf/psmi/shen2019}.
However, current studies mainly focus on segmentation (especially brain tumor segmentation), while the issue of missing modalities occurring in multi-modality cardiac analysis has not been well investigated.

There exist several studies that attempt to find alternative cardiac imaging modalities for missing modalities, namely modality replacement.
Modality replacement aims to find ``cheap" modalities (such as CTA and cine/ T1 mapping MRIs) to replace the ``expensive" ones (such as LGE MRI) by employing the inter-relationship among multi-modality images.
For example, the LV scarring areas presented in LGE MRI are correlated to LV Myo thicknesses measured by CTA \citep{journal/JCE/takigawa2019}. 
\citet{journal/Radiology/zhang2019} found that LV myocardial wall motion present in non-enhanced cine MRI could be used to predict chronic MyoI areas.
Therefore, one could employ CTA or cine MRI to detect scar regions instead of LGE MRI which is time-consuming to acquire and requires contrast agents  \citep{journal/FCM/o2021}.
Also, native T1 mapping MRI can be utilized to replace LGE MRI as it exhibits sensitivity to a variety of cardiac diseases \citep{journal/Cirir/zhang2021}.
So far, only a few modality replacement works \citep{journal/Cirir/zhang2021,journal/JCE/takigawa2019} have been proposed and are mainly devoted to LV Myo pathology analysis. 
It would be desirable to continue the study of modality relationships and develop modality replacement technologies for other cardiac substructure analyses.

\begin{figure*}[t]\center
    \subfigure[] {\includegraphics[width=0.49\textwidth]{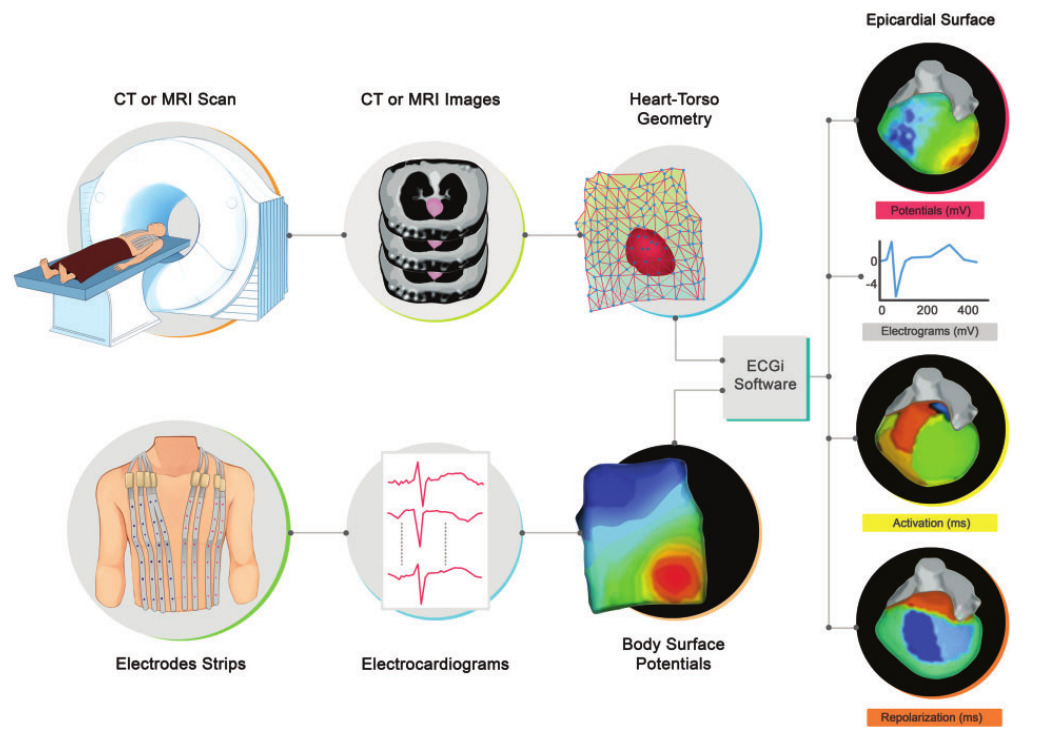}}
    \subfigure[] {\includegraphics[width=0.48\textwidth]{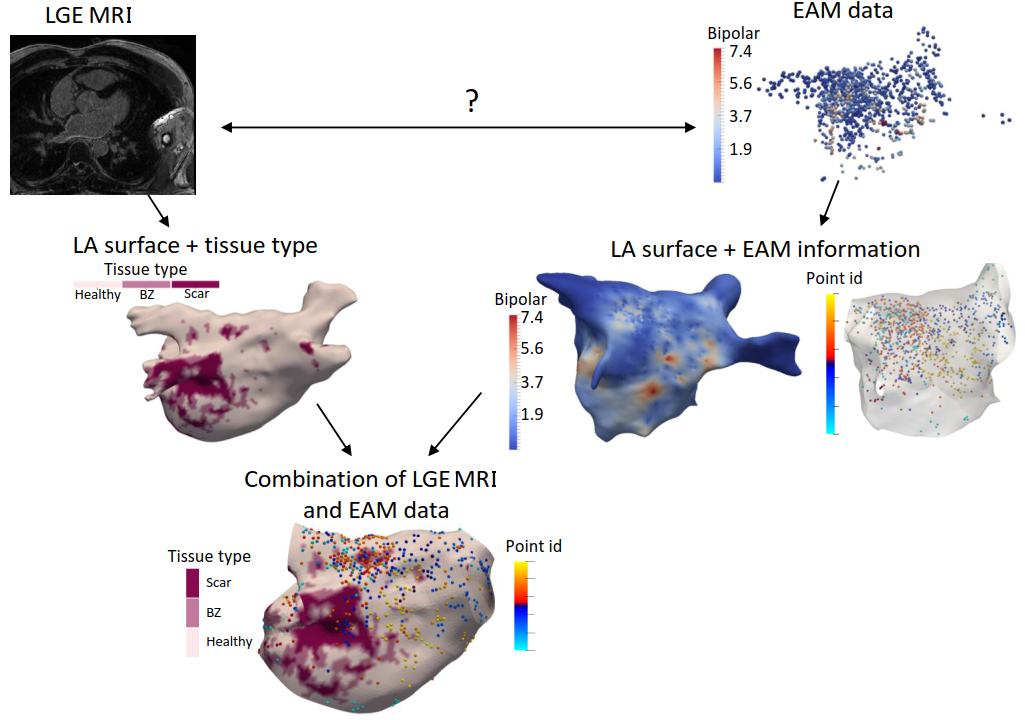}}
   \caption{ Two examples of combining cardiac imaging with non-imaging datasets.
   (a) The procedure of electrocardiographic imaging (ECGi), which combines CT/ MRI with ECG signal using mathematical algorithms.
  Here, body surface potentials are recorded from several electrodes, while the patient-specific heart-torso geometry is obtained from thoracic CT or MRI.
  Image adapted from \citet{journal/EP/pereira2020} with permission;
  (b) The procedure of combining LGE MRI and EAM data for LA scar localization and visualization. Image adapted from \citet{thesis/UPF/nunez2018} with permission.
   }
\label{fig:discussion:ECGI and EAM}
\end{figure*}

\subsection{Integration of cardiac imaging with non-imaging information} 

In addition to multi-modality images, there are many available non-imaging information sources associated with CVDs \citep{journal/NM/bai2020,journal/Nature/meyer2020}.
Take UK Biobank data as an example, which provides gender, age, BMI, biological information, heart and lung function measures, etc \citep{journal/Nature/bycroft2018}. 
Among these non-imaging modalities, ECGs can provide a substantial amount of information through its indirect recording of the electrical activity of the heart. 
This includes scar regions after MyoI, different arrhythmias, the effects of hypertension, as well as information about recovery and cardiac motion.
However, its interpretation requires considerable human expertise, and ECGs are limited in their ability to spatially locate and characterize CVDs.
The structural information from cardiac imaging data may be complementary to the information provided by ECGs. Body surface ECG mapping data, which provides a larger amount of measurements and includes more spatial information, may be preferable for this purpose \citep{journal/Rad/cochet2014,journal/EP/pereira2020}.
As presented in \Leireffig{fig:discussion:ECGI and EAM} (a), one could combine cardiac geometry obtained from CT/ MRI with body surface potentials for a more comprehensive noninvasive assessment of cardiac arrhythmias.
Specifically, the combined multi-modality information might be able to identify the mechanism of the arrhythmia and locate the circuit or source based on cardiac anatomy and myocardial substrate \citep{journal/Rad/cochet2014}.
It could then be embedded into a 3D mapping system to assist in catheter ablation therapy for atrial or ventricular arrhythmias.
Electroanatomical mapping (EAM) can also be integrated with cardiac imaging, such as CT \citep{journal/JICE/itoh2010}, MRI \citep{journal/JACC/reddy2004}, X-ray \citep{journal/Eur/scaglione2011}, to guide ablation procedures (see \Leireffig{fig:discussion:ECGI and EAM} (b)).
\citet{journal/JCR/wang2018} combined cine MRI with intra-ventricular pressure recording data to estimate diastolic myocardial stiffness and stress for personalized biomechanical analysis.
Non-imaging information can also be integrated to diagnose cardiac abnormalities, for example, combining ECG and phonocardiogram signals \citep{journal/PESM/chakir2020,journal/HR/li2020}.

The combination of imaging and non-imaging data will provide a transition from image space to patient-level information.
However, it could be more challenging than the integration of different imaging modalities.
This is because it brings a data format variation, resulting in the requirement of intermediate conversion steps for each data.
For example, cardiac imaging data usually needs to be converted into a surface, while ECG normally needs to be recorded on a body surface via a multi-electrode vest.
It is necessary to develop algorithmic techniques to learn high-level features from two types of data sources for a more elegant combination \citep{journal/EP/bacoyannis2021}.

\begin{figure}[t]\center
 \includegraphics[width=0.48\textwidth]{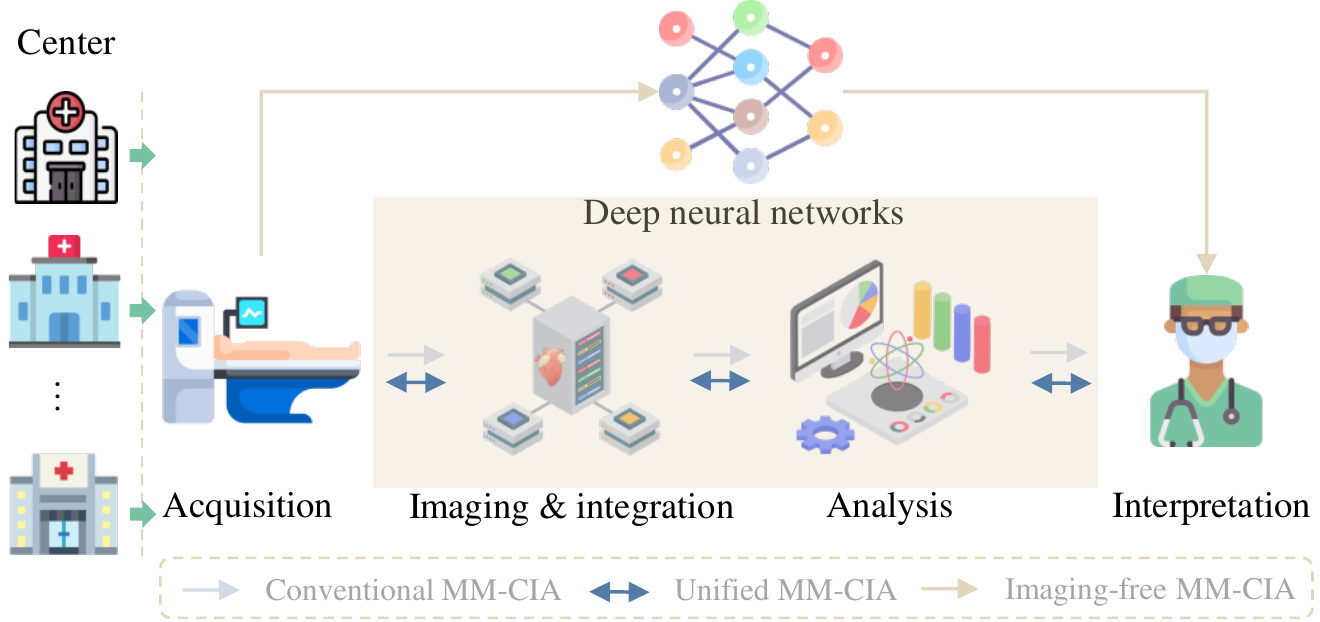}\\[-2ex]
  \caption{Traditional vs. future expected multi-center multi-modality cardiac image analysis (MM-CIA) procedures.
  Conventional MM-CIA is usually only optimized in a single direction and usually not for a clinical endpoint, while unified MM-CIA can transfer information in both directions.
  Imaging-free MM-CIA even can directly achieve CVD diagnosis from imaging acquisition in the absence of actual imaging.
  }
\label{fig:discussion:unified framework}
\end{figure}

\subsection{Unified framework for multi-center multi-modality cardiac image analysis} \label{unified and multi-center}
Current multi-modality image processing steps do not include much user interaction, resulting in the limited ability to adjust upstream processing steps based on downstream requirements.
Furthermore, these processing stages are not optimized for a clinical endpoint.
We therefore expect a unified framework for joint optimization of the whole pipeline with respect to a clinical endpoint, as shown in \Leireffig{fig:discussion:unified framework}.
For example, \citet{journal/TPAMI/pan2021} recently developed a joint image synthesis and disease diagnosis framework using incomplete multi-modality neuroimages.
This study emphasized the importance of performing a disease-image-oriented joint optimization for disease identification and prediction.

Due to different clinical acquisition protocols across hospitals, acquiring all modalities from all centers is costly and often impossible in clinical settings.
Federated learning (also referred to as collaborative learning) is a promising technique that can train models on multiple distributed centers that hold local data samples \citep{journal/SPM/li2020}.
It ensures data sharing while protecting patient data privacy, which would increase the number of available cardiac imaging with different modalities.
However, different centers may own misaligned imaging modalities, which raises a realistic challenge for federated learning.
Recently, \citet{journal/arXiv/chang2022} designed a privacy secure decentralized multi-modality adaptive learning architecture, namely
ModalityBank. 
ModalityBank consists of a domain-specific modulation parameters bank, a central generator, and multiple distributed discriminators located in a variety of medical centers.
It can synthesize multi-modality images by switching different sets of configurations to complete missing modalities across data centers.
Nevertheless, the domain generalization ability of current DL-based models needs to be significantly improved to handle multi-center multi-modality cardiac datasets \citep{conf/MICCAI/li2021}.
In the future, with deep neural networks we even could achieve direct CVD diagnosis from imaging acquisition without the requirement of imaging, which however could be too ambitious for now.

\begin{figure*}[t]\center
    \subfigure[] {\includegraphics[width=0.45\textwidth]{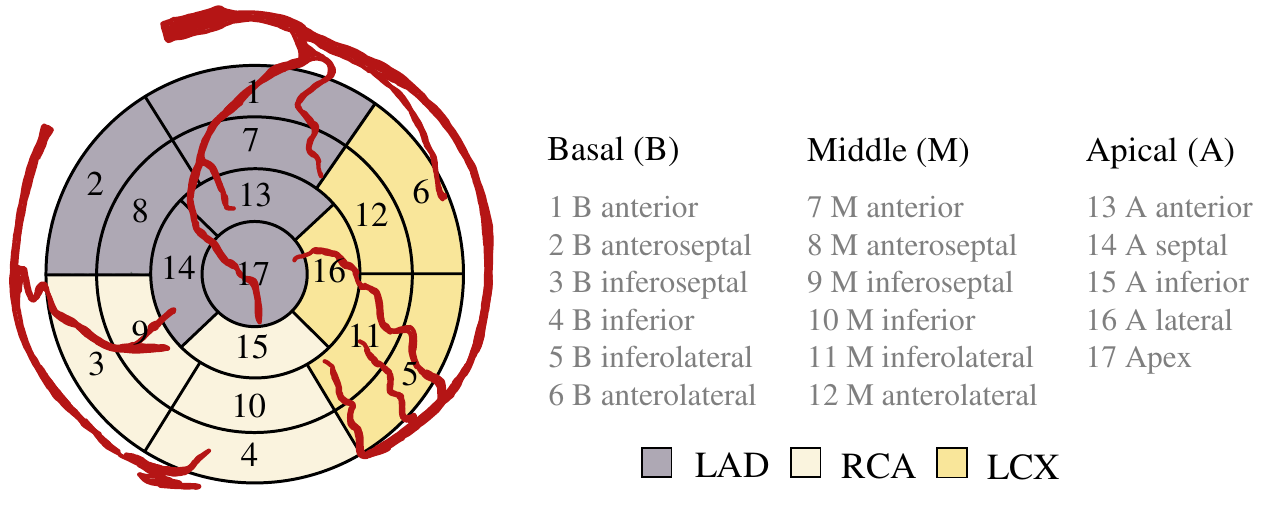}}
    \subfigure[] {\includegraphics[width=0.53\textwidth]{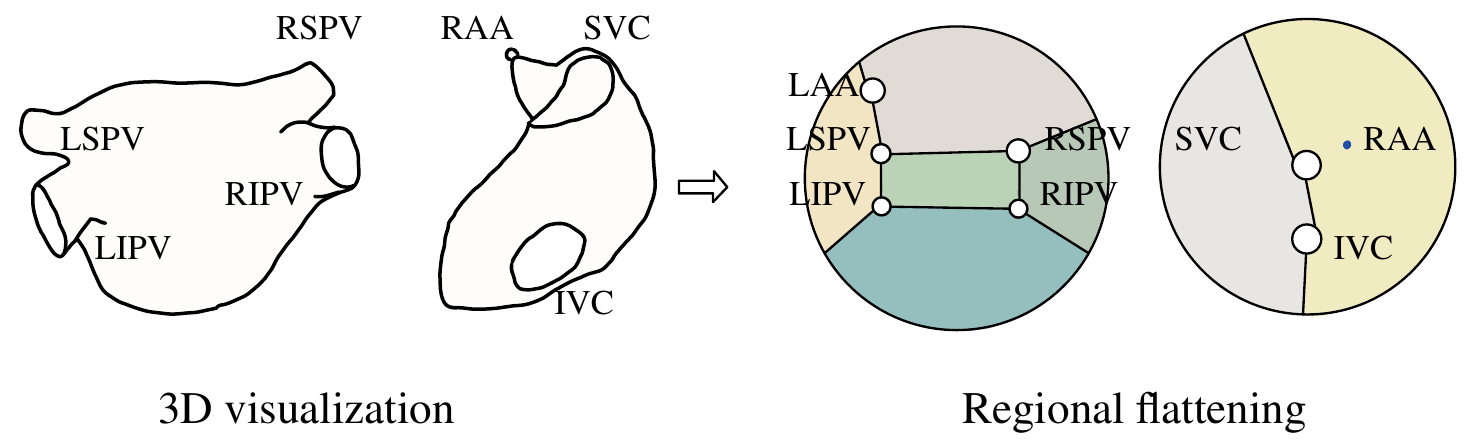}}
   \caption{
      (a) Illustration of AHA-17 bull's-eye plot and the CA regions with corresponding names (LAD: left anterior descending; RCA: right CA; LCX: left circumflex);
      (b) Regional flattening of the left atrium and right atrium. Image designed referring to \citet{journal/FIMH/nunez2019}.
   }
\label{fig:discussion:polar map}
\end{figure*}

\subsection{Polar map of multi-modality cardiac imaging} 
The visualization of multi-modality images requires a common coordinate system, where the spatial information about cardiac anatomy and physiology from different modalities can be combined.
One can utilize a unified 3D cardiac visualization, but it requires viewing rotated images or a rotating display.
In contrast, a 2D polar map can display the entire cardiac surface and the coronary anatomy in a single image, even though it comes at the expense of undesired geometric distortions.
The 2D polar map offers cardiac regional features, permits a comparison of spatial data across patients, and allows a standardized analysis across different imaging modalities.
Therefore, it is promising for multi-modality cardiac image analysis and could be useful in clinical practice.

The most popular application of polar maps to the heart is the shaped LV, which is often mapped into a planar bull's-eye plot \citep{journal/Cir/american2002}.
In the bull's-eye polar map, the LV can be divided into different segments (AHA-16, 17 or 18) for simplification and standardization \citep{journal/EHJCI/lang2015}.
\Leireffig{fig:discussion:polar map} (a) presents the most common AHA-17 map, which can be used to visualize regional function information about the LV, including wall motion, perfusion and distribution of scars and edema.
Bull's eye maps have also been applied to visualize the CA \citep{journal/JACC/nakahara2016,journal/JCAT/fukushima2020} and coronary sinus \citep{journal/IPCAI/ma2012}.
Specific to multi-modality image analysis, \citet{journal/JACC/nakahara2016} proposed a fusion based bull's eye map to describe the SPECT–coronary CTA data in a single image.
\citet{journal/TMI/tavard2014} employed a bull's eye map to visualize the fused anatomical, functional and electrical data acquired from CT, speckle tracking Echo, and electro-anatomical mappings (EAM).
\citet{journal/MedIA/paun2017,journal/MedIA/bayer2018,journal/MedIA/schuler2021} provided more general ventricle representations that included both LV and RV.

Compared to the ventricles, it is more challenging to generate a standard planar mapping for the atria which usually have more complex shapes \citep{journal/MedIA/roney2019,journal/MedIA/li2022b}. 
There are only a few studies targeting atrial polar mapping \citep{journal/JICE/williams2017,journal/MedIA/roney2019,journal/FIMH/nunez2019,journal/TVCG/nunez2020}.
Their templates were all designed for the most common LA topology with four pulmonary veins and are sensitive to LA topological variants.
\Leireffig{fig:discussion:polar map} (b) presents a universal atrial coordinate mapping system for 2D visualization of both the LA and right atrium.
For unified multi-modality image visualization and analysis, we expect more comprehensive and robust cardiac polar maps, which can cover more substructures of the heart and adapt for cardiac variations across patients.

\section{Conclusion} \label{conclusion}
The development of cardiac multi-modality imaging has already refined clinical decision-making and improved the treatment of patients with CVD.
Many studies have been proposed to integrate multiple imaging modalities, to facilitate our understanding of the complex anatomy of the heart and its behavior and a move to more patient-specific interpretation.
This review summarizes recent advances in multi-modality cardiac imaging techniques, computing algorithms, public datasets, evaluation measures, clinical applications and future perspectives. 
We conclude that although the development of machine learning, especially deep learning, has promoted multi-modality cardiac image analysis, there are still many unresolved issues.
Note that although the review focuses on the multi-modality cardiac image computing, the key issues and obstacles of the computing techniques can be generalized to other medical image analysis applications and utilized in mainstream clinical practice.

\section*{Acknowledgment}
This work was funded by the CompBioMed 2 Centre of Excellence in Computational Biomedicine (European Commission Horizon 2020 research and innovation programme, grant agreement No. 823712).
X. Zhuang was supported by the National Natural Science Foundation of China (grant no. 61971142, 62111530195 and 62011540404) and the development fund for Shanghai talents (no. 2020015).
L. Li was partially supported by the SJTU 2021 Outstanding Doctoral Graduate Development Scholarship.

\bibliographystyle{model2-names}
\biboptions{authoryear}
\bibliography{A_refs}

\end{document}